\documentclass[11pt,a4paper]{article}                           

\usepackage{a4}          % Remove before submission

\usepackage{amsmath}
\usepackage{amssymb}
\usepackage[usenames]{color} 
\usepackage{graphicx}
\usepackage{array}
\usepackage{hhline}
\usepackage{microtype}
\usepackage[bf]{caption}
\usepackage{color}
\usepackage[usenames,dvipsnames]{xcolor}
\usepackage{tcolorbox}
\usepackage{multirow}
\usepackage{cite}

\usepackage{ifpdf}
\ifpdf
  \usepackage[pdftex,
    pdftitle={},
    pdfauthor={},
    pdfsubject={},
    bookmarksopen, bookmarksnumbered, bookmarksopenlevel=2]{hyperref}
\fi
\def\hybrid{\topmargin -20pt    \oddsidemargin 0pt
        \headheight 0pt \headsep 0pt
        \textwidth 6.25in       % A4 paper
        \textheight 9 in       % A4 paper
        \marginparwidth .875in
        \parskip 5pt plus 1pt 
          \jot = 1.5ex
   }
\hybrid
\numberwithin{equation}{section}
\numberwithin{table}{section}

\newcommand{\beq}{\begin{equation}}  \newcommand{\eeq}{\end{equation}}
\newcommand{\bal}{\begin{aligned}}   \newcommand{\eal}{\end{aligned}}
\newcommand{\bea}{\begin{eqnarray}}  \newcommand{\eea}{\end{eqnarray}}

\def\ov{\overline}

\newcommand{\bmat}{\left(\begin{array}}
\newcommand{\emat}{\end{array}\right)}

\newcommand{\bbZ}{\mathbb{Z}}

\newcommand{\nn}{\nonumber}

\newcommand{\hD}{\hat D}
\newcommand{\cO}{\mathcal{O}}

\newcommand{\cC}{\mathcal{C}}

\newcommand{\cN}{\mathcal{N}}

\newcommand{\cF}{\mathcal{F}}

\newcommand{\PP}{\mathbb{P}}

\newcommand{\be}{\begin{equation}}
\newcommand{\ee}{\end{equation}}

\begin{document}

\baselineskip=14pt
\parskip 5pt plus 1pt

\vspace*{-1.5cm}
\begin{flushright}    % Publication numbers
  {\small
  }
\end{flushright}

\vspace{2cm}
\begin{center}        % Main title
  {\LARGE Supersymmetric Protection and the Swampland}
\end{center}

\vspace{0.5cm}
\begin{center}        % Authors
{\large  Eran Palti$^{1,2}$,\; Cumrun Vafa$^3$,\;  Timo Weigand$^{4}$}
\end{center}

\vspace{0.15cm}
\begin{center}        % Institutes
\emph{$^1$Max-Planck-Institut f\"ur Physik (Werner-Heisenberg-Institut), 80805 M{\"u}nchen, Germany}\\[.3cm]
\emph{$^2$Department of Physics, Ben-Gurion University of the Negev, Beer-Sheva 84105, Israel}\\[.3cm]
\emph{$^3$Jefferson Physical Laboratory, Harvard University, Cambridge, MA 02138, USA}\\[.3cm]
\emph{$^4$PRISMA Cluster of Excellence and Mainz Institute for Theoretical Physics, \\
Johannes Gutenberg-Universit\"at, 55099 Mainz, Germany}
             \\[0.15cm]
 
\end{center}

\vspace{2cm}

%%%%%%%%%%%%%%%%%%%%%%%%%%%%%%%%%%%%%%%%%%%%%%%
%%%%%%%%%%%%%%%%%%%%%%%%%%%%%%%%%%%%%%%%%%%%%%%
%%%%%%%%%%%%%%%%%%%%%%%%%%%%%%%%%%%%%%%%%%%%%%%
%%%%%%%%%%%%%%%%%%%%%%%%%%%%%%%%%%%%%%%%%%%%%%%
%%%%%%%%%%%%%%%%%%%%%%%%%%%%%%%%%%%%%%%%%%%%%%%
%%%%%%%%%%%%%%%%%%%%%%%%%%%%%%%%%%%%%%%%%%%%%%%
%%%%%%%%%%%%%%%%%%%%%%%%%%%%%%%%%%%%%%%%%%%%%%%
%%%%%%%%%%%%%%%%%%%%%%%%%%%%%%%%%%%%%%%%%%%%%%%

\begin{abstract}
\noindent   
For certain terms in the action, supersymmetry can forbid an infinite number of possible contributions. We study whether such protection can occur in quantum gravity even without sufficient supersymmetry. 
We focus on whether the superpotential can vanish exactly in four-dimensional $\cN=1$ theories, and if the prepotential can be exactly cubic in $\cN=2$ theories. 
We investigate these questions in string theory and find that for almost all known string constructions the corrections allowed by supersymmetry do occur. However, we do find some special settings where all the corrections can be proven to vanish.  These examples all share the common feature that they are related, through a certain orbifolding by a discrete gauged R-symmetry element, to a higher supersymmetric theory. Motivated by these results, we propose a Swampland criterion that any theory which enjoys such protection beyond its realised supersymmetry must have a direct connection to a higher supersymmetric theory. 
\end{abstract}

\thispagestyle{empty}
\clearpage

\tableofcontents

\setcounter{page}{1}

%%%%%%%%%%%%%%%%%%%%%%%%%%%%%%%%%%%%%%%%%%%%%%%
%%%%%%%%%%%%%%%%%%%%%%%%%%%%%%%%%%%%%%%%%%%%%%%
%%%%%%%%%%%                 %%%%%%%%%%%%%%%%%%%
%%%%%%%%%%%  DOCUMENT BODY  %%%%%%%%%%%%%%%%%%%
%%%%%%%%%%%                 %%%%%%%%%%%%%%%%%%%
%%%%%%%%%%%%%%%%%%%%%%%%%%%%%%%%%%%%%%%%%%%%%%%
%%%%%%%%%%%%%%%%%%%%%%%%%%%%%%%%%%%%%%%%%%%%%%%
%%%%%%%%%%%%%%%%%%%%%%%%%%%%%%%%%%%%%%%%%%%%%%%

%\newpage

%%%%%%%%%%%%%%%%%%%%%%%%%%%%%%%%%%%%%%%%%%%%%%%%%%%%%%%%%%%%%%%%%%%%%%%%%%%%
\section{Introduction}
\label{sec:intro}
%%%%%%%%%%%%%%%%%%%%%%%%%%%%%%%%%%%%%%%%%%%%%%%%%%%%%%%%%%%%%%%%%%%%%%%%%%%%

Supersymmetry is unique as a symmetry which relates external Poincar\'e charges such as mass to charges with respect to internal symmetries. Such relations are often critical
for the existence of certain inequalities between observables of the theory which are precisely saturated for distinguished states.
For example, BPS objects exhibit an exact equality between their mass and charge. Closely related to precisely saturated inequalities are precisely vanishing quantities. To name one, a theory which preserves $\cN=2$ supersymmetry in four dimensions has an exactly vanishing potential and can admit an exact moduli space. In supersymmetric theories there also exist quantities which enjoy partial protection. For example, the superpotential $W$ in four-dimensional theories with $\cN=1$ supersymmetry is allowed to be non-zero but has to be a holomorphic quantity in terms of chiral fields.  But can it be that $W=0$ and that quantum corrections do not contribute to $W$?  Similarly for $\cN=2$ theories gauge couplings are determined in terms of prepotentials which are generally not cubic (in flat coordinates) but receive corrections.  Could there be gravity theories with $\cN=2$ supersymmetry for which all such corrections vanish?  In the absence of gravity it is relatively straightforward to construct examples for which such corrections vanish.  But quantum gravity is typically far more restrictive as is well known in the context of the Swampland program \cite{Vafa:2005ui} (for reviews see e.g. \cite{Brennan:2017rbf,Palti:2019pca}). Whether or not the partially protected supersymmetric quantities can have additional vanishing properties in the quantum gravity context is therefore an excellent question for the Swampland program and consequently forms the topic of this paper.

We are particularly interested in whether there exist theories in quantum gravity without the sufficient amount of supersymmetry to guarantee the absence of certain non-perturbative corrections but which nonetheless receive no such corrections. We will see in examples that all quantities which can be corrected by quantum effects are corrected unless the theory is related, in a subtle way, to one with higher supersymmetry responsible for their protection.
Let us introduce the quantities we will consider, restricting ourselves to four-dimensional theories throughout this paper.

\subsubsection*{$\cN=1$ supersymmetry}

In $\cN=1$ supergravities the superpotential $W$ is well-known to receive no perturbative quantum corrections due to holomorphicity in the couplings combined with shift symmetries, and the gauge kinetic function $f$ receives only 1-loop perturbative corrections. However, they both are subject to potentially an infinite number of non-perturbative corrections. In particular, if we suppose that the classical superpotential $W_{\rm cl.}\left(\Phi\right)=0$, the only allowed contributions to the full superpotential take the form
\be
W =  \sum_{n,i} A_{n}\left(\Phi\right) e^{-a^i_n \Phi_i} \;.
\ee
Here $n$ labels the different instanton contributions, while $i$ runs over the chiral superfields.\footnote{The exponential terms may also be generated by gaugino condensation in the infrared, but we will focus on the instanton contributions.} The $A_n$ are (potentially vanishing) holomorphic functions of the fields while the $a_n^i$ are constants. Similarly, the gauge kinetic function has an analogous holomorphic expansion, but with an additional potential 1-loop contribution. The question of interest is whether it is possible to have 
\be 
A_{n}=0 \;\;\mathrm{for\;all\;} n \;.
\label{an0}
\ee 
Note that in principle one can split the condition (\ref{an0}) into two cases, according to whether it is satisfied for all values of the fields $\Phi_i$, or whether there is only a sub-locus in the space of all the massless fields where $W=0$. In $\cN=2$ vacua (\ref{an0}) is automatically satisfied all over field space. 

\subsubsection*{$\cN=2$ supersymmetry}

In $\cN=2$ theories, for instance for Type IIA string theory compactified on a Calabi-Yau three-fold, the prepotential $\cal F$ of the vector multiplet sector contains a term cubic in the superfields, lower polynomial terms and exponential terms,
\be
\cF = \cF_{\mathrm{polynomial}}\left(\Phi\right) + \sum_{n,i} B_{n}\left(\Phi\right) e^{-a^i_n \Phi_i} \;,
\ee
where the exponential terms are generated by worldsheet instanton effects.
We may ask if the infinite series of exponential terms are always present or whether it is possible to have
\be 
B_{n}=0 \;\;\mathrm{for\;all\;} n \;.
\label{bn0}
\ee 
In $\cN=4$ vacua, such as for Type II compactifications on $K3 \times {\mathbb T}^2$, (\ref{bn0}) is automatically satisfied.

We will refer to the quantities like those appearing on the lefthand side of (\ref{an0}) and (\ref{bn0}) as  {\it supersymmetric protected quantities} (SPQs) in the following sense: They are of restricted form, but expected to be non-vanishing for {\it generic} theories with $\cN=1$ or $\cN=2$ supersymmetry, respectively, while they vanish in presence of higher supersymmetry.
 Forbidding an infinite number of operators, without any symmetry responsible for it, would require an infinite amount of accidental cancellations. 
If there were an infinite number of consistent quantum gravity theories, such accidental cancellations might indeed occur in concrete theories.
On the other hand, if the number of consistent theories of quantum gravity is finite, as is widely believed, such a cancellation is extremely unlikely.
Invoking a principle of genericity, it is therefore natural to expect that whenever a protected quantity is allowed by supersymmetry, it is non-zero unless it is protected (in some way) by a symmetry.

The goal of this paper is to sharpen this natural expectation by constructing explicit examples of gravitational theories in which the supersymmetric protected quantities vanish. 
Our findings motivate us to propose a \\
% In particular, we will sharpen the way in which a discrete gauge symmetry can protect a SPQ, by proposing that it allows a certain sub-sector to feel an enhanced supersymmetry. More precisely we propose the following \\
\newline
{
{\bf Supersymmetric Genericity Conjecture:} {\it A theory of quantum gravity 
where a supersymmetric protected quantity (as defined above) vanishes, even though this is not required by the amount of supersymmetry preserved by the theory, must be related to a higher supersymmetric theory.} \\

Note that this conjecture is in line with the general principle that anything which is generically allowed in quantum gravity is `enforced' and hard to prevent and that there occur no accidental vanishings.  In the context of quantum gravity theories, for which the finiteness of the allowed possibilities is a Swampland principle, this is plausible: We cannot fine-tune parameters to avoid the generic prediction due to the finiteness of the number of possibilities.

All the examples of quantum gravity theories with vanishing supersymmetric protected quantities as constructed in this paper 
have the following characteristics:
There exists a discrete gauge symmetry $\tilde \Gamma$ whose neutral sector is identical to the neutral sector of a higher supersymmetric theory under a discrete R-symmetry $\Gamma$.\footnote{Note that, unlike $\Gamma$, the discrete gauge symmetry $\tilde \Gamma$ is not an R-symmetry. In all examples studied in this work, $\Gamma$ is abelian and $\tilde \Gamma$ and $\Gamma$ are isomorphic as groups. } 
Furthermore, if all {\it massless} fields in the theory are neutral under $\tilde \Gamma$, then the supersymmetric protected quantities vanish everywhere in moduli space. 

The examples studied in this paper are in agreement with these characteristics being necessary conditions for the appearance of additional protection beyond the amount of supersymmetry realized in a theory.
As we will further exemplify, however, they are not sufficient. 
Also, while we do not have any explicit realizations, we do not rule out other possible connections to higher supersymmetric theories.

}

Our guiding principle will be explicit constructions in string theory.
Determining whether a supersymmetric protected quantity vanishes is particularly delicate in the case of $\cN=1$ supersymmetry. As a first step we will revisit stringy $\cN=1$ theories for which the superpotential corrections (\ref{an0}) have been proposed to vanish in the literature even though there is no symmetry argument  apparent to us which would be responsible for this. We will identify in all such cases subtle effects that have not been accounted for properly and  which in fact suggest that the superpotential is non-vanishing. At a technical level, one such effect is the lifting of certain deformation zero-modes for D3-brane instantons in F-theory by quartic fermionic terms in the instanton effective action, as well as the role of stringy spacetime instantons in the heterotic duality frame.
Based on this we will construct compactifications free of these subtleties and extract their common properties. We will show that they do satisfy the Supersymmetric Genericity Conjecture, and
in fact exhibit the characteristic properties detailed above.
Note that this conjecture, if correct, in particular states that exact four-dimensional supersymmetric Minkowski vacua can occur in quantum gravity only in special settings enjoying a protection mechanism which can be traced to some underlying higher supersymmetry.

The examples of theories which we study in this paper fall into two classes, corresponding to whether the supersymmetric protected quantity (SPQ) vanishes for all values of the massless fields  - in this case there are no massless fields charged under $\tilde \Gamma$ - 
or only on the sub-locus in field space where the massless fields charged under $\tilde \Gamma$ take a zero value.
Technically, {\it one} way to understand the vanishing of the SPQ is to note that in both cases there exists some duality frame where all the instantons which could create the SPQs exhibit too many fermionic zero-modes.
The zero-modes in question are associated with the breaking of a higher supersymmetry by the instanton. They can couple to the massless fields charged under $\tilde \Gamma$ and can hence be saturated only if the latter receive a non-zero vacuum expectation value. 
This explains why in absence of such massless charged fields the SPQ vanishes everywhere in moduli space.

The paper is set out as follows. In section \ref{sec:strpotcor} we discuss compactifications of string theory which may naively appear to have vanishing SPQs and show that such a conclusion is not justified due to non-perturbative effects that have not been taken into consideration. In section \ref{sec:strnocorr} we then study string theory compactifications which are free from the subtleties due to such effects. In section \ref{sec:charexam} we discuss the common features of these theories and propose a Swampland conjecture based on these features. We discuss our results in section \ref{sec:dis}. In the Appendices we present various technical details as well as further supporting evidence 
for the picture advocated in this work.

%%%%%%%%%%%%%%%%%%%%%%%%%%%%%%%%%%%%%%%%%%%%%%%%%%%%%%%%%%%%%%%%%%%%%%%%%%%%%
\section{String compactifications with generic corrections}
\label{sec:strpotcor}
%%%%%%%%%%%%%%%%%%%%%%%%%%%%%%%%%%%%%%%%%%%%%%%%%%%%%%%%%%%%%%%%%%%%%%%%%%%%%

In this section we present a study of non-perturbative corrections to partially protected quantities in string theory. 
According to the discussion in the introduction, such corrections are expected to occur unless they are disallowed by a certain type of symmetry.
This is in seeming contradiction with candidates reported in the literature for string compactifications without  an obvious protection mechanism, but which nonetheless lack, for instance, an $\cN=1$ superpotential.
% We focus on such settings which were expected to have no such corrections, and 
We will exemplify that in these cases unaccounted instanton effects can spoil the protection, in agreement with our expectations.

\begin{figure}[t]
\centering
 \includegraphics[width=0.8\textwidth]{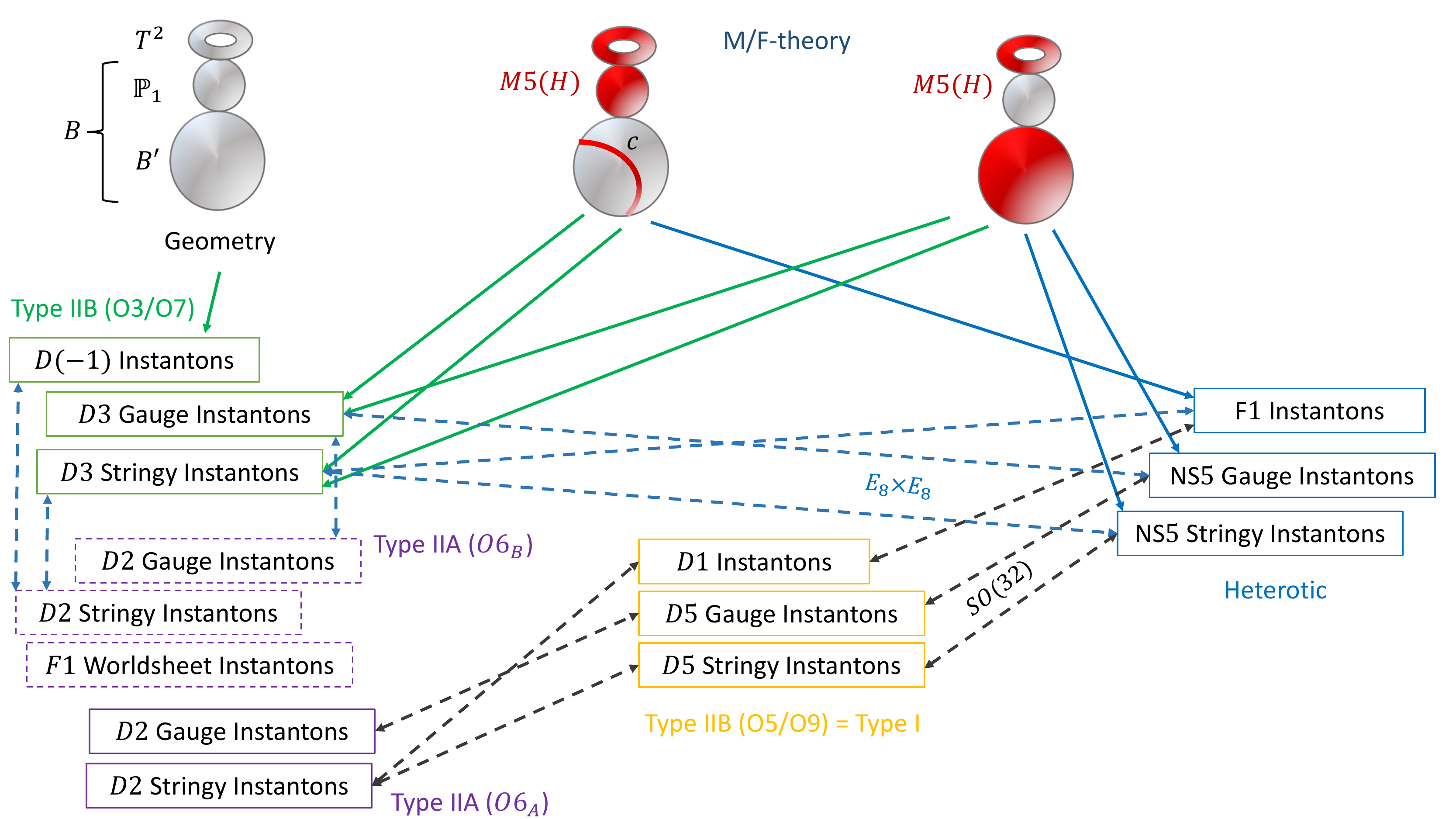}
\caption{Various types of instantons that appear in string theory, and their relations through dualities given by dashed lines. The dualities can be understood through their M-theory origin, which can also be used to calculate the (neutral) zero-modes of the instantons.}
\label{fig:instmap}
\end{figure}

String theory offers a unique window into non-perturbative quantum gravity effects. Specifically, instanton corrections can be understood in terms of extended (non-perturbative) objects in the theory wrapping cycles in the extra dimensions. 
%In some cases we know the microscopic physics of these branes, which gives a way to calculate the zero-modes of the instantons. This is the crucial factor for determining if they contribute to the partially protected holomorphic quantities. 
Some of the brane instantons can be understood as gauge theory instantons \cite{Witten:1995gx} (see e.g. \cite{Bianchi:2007ft} for a review), while most instantons have no such gauge theory interpretation and are called stringy instantons. 
The different types of instantons and their relations through dualities, particularly within the duality orbit involving M-theory and F-theory, are shown in Figure \ref{fig:instmap}. Many aspects of stringy instantons relevant to the following analysis, especially the counting of instanton zero-modes, can be found e.g. in the review \cite{Blumenhagen:2009qh} and references therein.

The work reported in this section involves a significant amount of technical details, as well as new effects which are important for instantons. For example, we prove for the first time the existence of stringy spacetime instantons effects in the heterotic string (which have no gauge theory interpretation) which correct the $\cN=1$ superpotential, and propose that certain instanton zero-modes for D-brane instantons in Type II string theory/F-theory can be lifted by quartic interaction terms in presence of suitable spacetime filling branes. Some details and technicalities are presented in the Appendices. %and we are brief in reporting the results in the section. 

%%%%%%%%%%%%%%%%%%%%%%%%%%%%%%%%%%%%%%%%%%%%%%%%%%%%%%%%%%%%%%%%%%%%%%%%%%%%%
\subsection{F-theory vacua with $\cN=1$ supersymmetry}
\label{sec:ftvac}

F-theory presents possibly the best understood setting for studying non-perturbative effects \cite{Witten:1996bn}, through its duality with M-theory. Instantons which appear to be at very different footing in other duality frames are all understood as M5-branes wrapping divisors in Calabi-Yau fourfolds. A generic Calabi-Yau is expected to support many appropriate divisors for instantons to contribute to the superpotential. In order for this to happen, the instanton must not exhibit any fermionic zero-modes modes besides the two universally present modes $\theta^\alpha$ unless they can be saturated in the path integral due to suitable interactions.

%For F-theory on an elliptically fibered Calabi-Yau 4-fold,

 The additional zero-modes, if present, fall into two classes: Extra Goldstino zero-modes are related to the breaking of a higher supersymmetry by the instanton. %and are the subject of section \ref{sec:genconst}. 
 Their presence for {\it every} instanton in the theory requires a non-generic structure of the theory for which we will construct examples in Section \ref{sec:strnocorr}.
 All remaining zero-modes admit no such interpretation. Their presence is hence accidental, from the perspective of supersymmetry, and
 according to the logic proposed in the Introduction, there is no reason why {\it every} instanton divisor should exhibit them, in generic situations.

In this section, we will further corroborate this reasoning by re-examining the status of  the non-Goldstino zero-modes.
%The presence of the extra universal modes for {\it all} instantons, on the other hand, will be identified, in section, as due to a certain discrete gauge symmetry which is related to an R-symmetry of a higher supersymmetric theory.
The question of whether there are special examples which receive no instanton corrections due to such modes was already preliminarily addressed in the pioneering work \cite{Witten:1996bn}. It was proposed that there are certain simple
F-theory backgrounds, for example $\mathbb{P}^3$ and $\mathbb{P}^1 \times \mathbb{P}^2$, which admit no appropriate divisors for the instantons to contribute to the superpotential. However, further investigations, in particular \cite{Bianchi:2011qh}, and new results in this paper show that these examples can, and sometimes certainly do, receive instanton corrections. 

\subsubsection{Instantons on generic Calabi-Yau fourfolds}
\label{sec:instgency}

Let us first recall the geometric criterion of \cite{Witten:1996bn} for instantons to contribute to the superpotential in F-theory. Consider a Calabi-Yau $Y_4$  which is elliptically fibered over $B_3$,
\be
\pi: Y_4 \rightarrow B_3 \,.
\ee
Under F/M-theory duality a D3-brane instanton in F-theory (oftentimes referred to as an {\it E3-brane}), along a divisor $D \subset B_3$, dualizes to an M5-brane instanton in M-theory wrapping the associated vertical divisor
\be \label{verticaldivisor}
{\hD} = \pi^*(D) 
\ee 
on $Y_4$. In addition to specifying the divisor $D$, one must sum over all admissible instanton flux configurations, which correspond to suitable 3-form fluxes on $\hD$ \cite{Kerstan:2012cy} or 2-form fluxes on $D$ \cite{Grimm:2011dj,Bianchi:2011qh}. 

A {\it sufficient} criterion for an instanton along $\hD$ with zero flux to contribute to the superpotential can be stated \cite{Witten:1996bn} as
\be \label{h00vanish}
h^{0,0}(\hD)=1 \,, \qquad h^{i,0}(\hD)=0 \,, \qquad i=2,3,4 \,,
\ee
where the $h^{i,0}(\hD)$ count the various types of instanton zero-modes as summarized in Table \ref{tab:instmap-main}.
The fermionic zero-modes $\theta^\alpha$ (with superpartner $x^\mu)$ are universally present.
As will be discussed in section \ref{sec:genconst}, the modes $\bar\tau^{\dot\alpha}$ play the role of Goldstino modes for the breaking of a higher supersymmetry felt locally by the instanton, but in generic situations there is no rationale for them to be present for every instanton.
% and $\bar\tau^{\dot\alpha}$ are distinguished as universal zero-modes and play the role of Goldstinis for the breaking of part of the supersymmetry by the instanton.
Likewise, the remaining zero-modes are not protected by any form of supersymmetry except in special situations such as the ones analyzed in section \ref{sec:strnocorr}.
Generically they are simply associated with  internal Wilson line or geometric deformation degrees of freedom of the instanton divisor.\footnote{Apart form these uncharged instanton zero-modes, instantons can carry zero-modes which are charged under the gauge group of the effective theory.
For D3-branes instantons these are due to zero-modes in the sector of open strings between the instanton and the spacetime filling branes \cite{Ganor:1996pe}.
For stringy instantons not related to gauge instantons, saturating these zero-modes induces an operator involving charged fields in the instanton contribution \cite{Blumenhagen:2006xt,Ibanez:2006da,Florea:2006si,Haack:2006cy}.
For more details we refer e.g. to  \cite{Blumenhagen:2009qh} and references therein.
}

If (\ref{h00vanish}) holds, the arithmetic genus 
\be
\chi(\hat{D}) \equiv \sum_{i=0}^4 (-1)^i h^{i,0}(\hat{D}) 
\ee
takes the value
\be
\chi(\hD)=1 \;.
\label{chi1con}
\ee
The unit value of $\chi(\hD)$ is neither a sufficient nor necessary condition for an instanton to contribute to the superpotential, but it is a crude first order test. One advantage of it is that it is topological and can be easily computed as
\be \label{agen}
\chi(\hD)=  -\frac{1}{24}\int_{Y_4} (\hD^4 + c_2(Y_4) \hD^2)  = -\frac{1}{24} \int_{Y_4} c_2(Y_4) \hD^2 \,,
\ee
where the second equality uses the fact that $\int_{Y_4} \hD^4 =0$ for a vertical divisor.
Hence \cite{Klemm:1996ts}
\be
\chi(\hD) =1 \iff  \int_{Y_4} c_2\left(Y_4\right) \hD^2 = -24 \,.
\label{chitopcond}
\ee
This formulation gives a rough first estimate of when we might expect a given Calabi-Yau $Y_4$ to contain a divisor appropriate to host an instanton. 

To see that generically one expects many such divisors, let the cone of effective divisors on $Y_4$ be spanned by the set $\cC_\alpha$ and write $\hD = \sum_{\alpha} p^{\alpha} \cC_{\alpha}$ with $p^{\alpha}$ arbitrary positive integers. The intersection numbers of the $\cC_{\alpha}$, as appearing in (\ref{chitopcond}), will be some order one integers. Therefore, (\ref{chitopcond}) is a single condition on many free integers, which is generically expected to have many possible solutions.\footnote{The complication is of course to find solutions over the integers.} However, there are some exceptions to this admittedly crude expectation, to which we now turn. 

\begin{table}[t!]
  \centering
  \begin{tabular}{c|c|c|c|c}
    zero-modes                    &  statistics  & Type IIB   &   F-theory       & M-theory      \\ 
    \hline\hline
    $(X_\mu,\theta_\alpha)$       & (bose, fermi) & $h_+^{0,0}(\tilde D)$ & $h^{0}(D,{\cal O})$ & $h^{0,0}(\hD)$ \\  
    \hline
    $\ov\tau_{\dot\alpha}$        & fermi        & $h_-^{0,0}(\tilde D)$ &$h^{0}(D,K_{B_3}|_D)$    & \multirow{2}{*}{$h^{1,0}(\hD)$} \\
    $\gamma_{\alpha}$             & fermi        & $h_+^{1,0}(\tilde D)$ &   $h^{1}(D,{\cal O})$  &  \\
    \hline
    $(w, \ov\gamma_{\dot\alpha})$ & (bose, fermi) & $h_-^{1,0}(\tilde D)$ &$h^{1}(D,K_{B_3}|_D)$  & \multirow{2}{*}{$h^{2,0}(\hD)$} \\
    $\chi_{\alpha}$               & fermi & $h_+^{2,0}(\tilde D)$ &$h^{2}(D,{\cal O})$   & \\
    \hline
    $(c , \ov\chi_{\dot\alpha} )$ & (bose, fermi)        & $h_-^{2,0}(\tilde D)$ & $h^{2}(D,K_{B_3}|_D)$  & $h^{3,0}(\hD)$
  \end{tabular}
  \caption{\small Type IIB and F-theory zero-modes for instantons carrying no instanton flux, as modified from \cite{Blumenhagen:2010ja}. The M5-brane is wrapping the divisor $\hD$ on an elliptically fibered Calabi-Yau 4-fold $Y_4$ with base $B_3$ and canonical bundle $K_{B_3}$, while the Type IIB D3 instanton is wrapping the divisor $\tilde D$ on the Type IIB Calabi-Yau 3-fold $X_3$; $\tilde D$ is the double cover of the divisor $D$ given by the projection of $\hat D$ to $B_3$. The Type IIB counting applies to so-called $O(1)$ instantons, for which $\tilde D$ is invariant under the orientifold involution as a divisor, but not pointwise. The last three columns give the homology classes counting the respective types of modes in absence of instanton flux. In the Type IIB case, the homology is split into orientifold even and orientifold odd sectors \cite{Blumenhagen:2010ja}. The modes counted by $h_\pm^{2,0}(\tilde D)$ may be lifted by instanton flux \cite{Bianchi:2011qh}. The counting in the fourth column follows from a Leray sequence argument and is valid only if $\hat D$ is smooth \cite{Grassi:1997mr}.}
  \label{tab:instmap-main}
\end{table}

%%%%%%%%%%%%%%%%%%%%%%%%%%%%%%%%%%%%%%%%%%%%%%%%%
\subsubsection{Deformation zero mode lifting by instanton flux} 
\label{sec_P3}

An example of (an elliptically fibered) Calabi-Yau four-fold $Y_4$ which supports no instanton divisors satisfying the condition (\ref{chi1con}) was presented in \cite{Witten:1996bn}.
It is given by a generic Weierstrass model over the base $B_3 = \mathbb P^3$.
Such fibration can be realized  for instance as a hypersurface in an appropriate $\mathbb{P}^2$ bundle over a base $\mathbb{P}^3$. 
M5-brane instantons with a potential contribution to the superpotential in F-theory wrap a divisor $\hD = \pi^*D$, where $D  = a \, H$ is a positive integral multiple of the hyperplane class on $\mathbb P^3$.
The cohomology groups can be computed explicitly as \cite{Grassi1991,Witten:1996bn} 
\be
h^{i,0}(\hD) = (1,0,N_{\PP^3}(a-4),N_{\PP^3}(a)) \,,
%\h^{0,0}(\hD) =  1\;,\;\; h^{1,0}(\hD) =  0 \;,\;\;  h^{2,0}(\hD) = N_{\PP^3}(a-4) \;,\;\; h^{3,0}(\hD) = N_{\PP^3}(a) \;,
\ee
where $N_{\PP^3}(x)$ denotes the number of polynomials of homogenous degree $x$ in $\PP^3$. Note that all the additional zero-modes are due to zero-modes other than the Goldstino modes of type $\bar \tau$.

It is clear that no divisor $\hD$  with $h^{i,0}(\hD) = (1,0,0,0)$ exists, and in particular $\chi(\hD) \neq 1$ for any $a > 0$. 
It was therefore proposed in \cite{Witten:1996bn} that for such a Calabi-Yau no non-perturbative superpotential is generated.

However, there are two problems with such a conclusion. The first is related to the presence of spacetime-filling D3 branes, necessary to cancel the tadpoles, and will be exemplified in a different setting in section \ref{sec_P1P2}. The second problem is that this conclusion only takes into account instantons which support no instanton flux, and such fluxes can modify the zero mode counting. Indeed, in \cite{Bianchi:2011qh} it was shown that the inclusion of instanton flux leads to the generation of a superpotential, as we will review in Appendix \ref{appendix_Flux}. 
For example the degree one divisor $a=1$ on $\mathbb P^3$ corresponds to $D = \PP^2$ with $h^{2,0}(\hD)=0$ and $h^{3,0}(\hD)=3$. 
As discussed in detail in \cite{Bianchi:2011qh} by invoking a weakly coupled orientifold limit, the double cover $\tilde D$ of this divisor in the associated Type IIB Calabi-Yau $X_3$ allows for instanton flux which lifts all three deformation zero-modes. Hence a non-perturbative superpotential is indeed generated.

%%%%%%%%%%%%%%%%%%%%%%%%%%%%%%%%%%%%%%%%%%%%%%%%%
\subsubsection{Deformation zero mode lifting by D3 brane interactions}
\label{sec_P1P2}

For the case $B_3 = \mathbb P^1 \times \mathbb P^2$, it is still true \cite{Witten:1996bn} that there are no divisors which satisfy (\ref{chi1con}). However, this time it is not possible to lift the additional zero-modes with instanton flux using the mechanism of \cite{Bianchi:2011qh}. We show this in Appendix \ref{sec:nofluxlif}. 

F-theory on $B_3 = \mathbb P^1 \times \mathbb P^2$ (with vanishing background flux) requires the introduction of
\be
N_{D3} = \frac{1}{24} \chi(Y_4) = 822
\ee
spacetime-filling D3-branes to cancel the tadpole. 
We now argue that in the presence of these D3 branes subtle effects can make instantons, even with certain extra zero-modes, contribute to the superpotential. The key point is that there may be certain 4-fermion terms in the instanton action which can saturate the zero mode integral.  
This effect can saturate the zero-modes counted by $H^{3,0}(\hat D)$. As we show in Appendix  \ref{sec:nofluxlif}, there do exist divisors on $B_3 = \mathbb P^1 \times \mathbb P^2$ whose only additional zero-modes are of this type, for instance the instanton with zero mode structure $h^{i,0}(\hat D) = (1,0,0,2)$ in equ. (\ref{pHinstanton}).
This suggests the existence of a D3-brane instanton generated superpotential even in this geometry.\footnote{Note that the effect we are about to describe is different from the well-known interaction of spacetime-filling D3-branes with D3-instanton zero-modes first found in \cite{Ganor:1996pe}: If the E3-brane already has the correct number of (neutral) zero-modes to contribute to the superpotential, the Pfaffian prefactor of the instanton hits a zero when the instanton intersects the D3-branes. This is due to additional zero-modes between the instanton and the D3 brane. Moving the D3 brane away from the instanton makes these additional strings massive and removes the zero-modes. In our case, the D3-E3 zero-modes participate in quartic couplings involving extra E3-brane deformation modes, which lift both types of zero-modes.}

It is informative to study this effect in the Type IIB limit where we can have control over the microscopic physics of the instanton. The map from the F-theory zero-modes to the Type IIB zero-modes is shown in Table \ref{tab:instmap-main}. 

Consider the anti-chiral modes $\bar\chi^{\dot \alpha}$ counted by $h^{3,0}(\hat D) = h^{2,0}_-(\tilde D)$. These are always paired with a complex bosonic mode $c$ representing a geometric deformation of the instanton cycle $\tilde D$ within the Type IIB Calabi-Yau $X_3$.
In order for the instanton to contribute to the superpotential, it must be possible to saturate the integral 
\be
\int d^4x \, d^2\theta \, d^2\bar{\chi} \, d^2 c \,\,  e^{-S_{inst}} \;.
\ee
The Grassmann measure $d^2\bar{\chi}$ can be saturated through an interaction term in the instanton action of the form $S_{inst}^{int} = \bar{\chi} \, \cO \, \bar{\chi}$, with $\cO$ an appropriate operator. Suppose that the D3-brane and the instanton are coincident. In such a situation there exist open-string zero-modes at the instanton-brane intersection, which we denote by Grassmann-valued scalars $\lambda_{DE}$ and ${\lambda}_{ED}$, with subscripts denoting the start and end point of the open string. 
Then provided there exists a (tree-level) interaction term
\be \label{Sintquartic-main}
S_{inst}^{int} \sim \lambda_{DE} \, \bar{\chi}^{\dot \alpha} \, \bar{\chi}_{\dot\alpha} \, {\lambda}_{ED}\;,
\ee
 the zero mode integral $\int d^4x \, d^2\theta \, d^2\bar{\chi} \, d\lambda_{DE} \, d\lambda_{ED} \, e^{-S_{inst}}$ is saturated. 
 Note that this quartic term is accompanied by a cubic coupling 
 \be
 \lambda_{DE} \, c \, \lambda_{DE} \;,
 \ee
 involving the bosonic instanton deformation mode $c$.

From the perspective of Type IIB string theory, the interaction term (\ref{Sintquartic-main}) corresponds to the zero (external) momentum limit $k \rightarrow 0$ of the disc diagram involving four vertex operator insertions 
\be \label{Discchi-main}
S_{inst}^{int}  \sim \lim_{k\rightarrow 0}  \left<V^{-1/2}_{\lambda_{DE}}V^{-1/2}_{\bar\chi}V^{-1/2}_{\bar\chi}V^{-1/2}_{\lambda_{ED}}\right>_{Disc} \,.
\ee
The superscripts on the vertex operators give the ghost charge (or picture), which for the disc must sum to $-2$. Even without calculating (\ref{Discchi-main}) explicitly, we note that it is allowed by all the worldsheet selection rules. 
This follows from the fact that  the total $U(1)$ worldsheet charges of the different operators are given by
\be \label{totalU1vertexop}
V^{-1/2}_{\bar\chi}: \, \, Q = \frac{1}{2} \,, \qquad \quad V^{-1/2}_{\lambda_{DE}/\lambda_{ED}}: \, \, Q =  - \frac{1}{2} 
\ee
and hence cancel in the correlator (\ref{Discchi-main}). While this does not yet show that the limit of vanishing momentum $k \to 0$ leads to a contact term in  (\ref{Discchi-main}), at least the correlator is consistent with the worldsheet selection rules.

To go beyond this statement requires performing an explicit CFT computation of the correlator, which is in principle possible for instance
in a toroidal setting where $X_3 = {\mathbb T}^2_1 \times {\mathbb T}^2_2 \times {\mathbb T}^2_3$ (or an orbifold thereof) and where we can construct the vertex operators explicitly. 
 Associated with each of the three two-tori ${\mathbb T}^2_i$ there is a  $U(1)_i$ selection rule on the worldsheet, the sum of which corresponds to the above total $U(1)$ worldsheet charge present on generic Calabi-Yau three-folds. 
 For an E3-brane wrapping, e.g., along ${\mathbb T}^2_2 \times {\mathbb T}^2_3$,
  the individual charges are given by (see for example \cite{Cvetic:2007ku})
\be
V^{-1/2}_{\bar\chi}: Q_i = \left(-\frac{1}{2},\frac{1}{2},\frac{1}{2}\right)  \,, \qquad \quad V^{-1/2}_{\lambda_{DE}/\lambda_{ED}}: Q_i =  \left(- \frac{1}{2},0,0\right),
\ee
leading to a collection of worldsheet charges for the correlator (\ref{Discchi-main}) of
 $\left(-2,1,1\right)$. Therefore, the operator is forbidden by these individual selection rules.  This is to be contrasted with the above smooth $\cN=1$ background, where the isometries associated to the $U(1)$s are broken and the total worldsheet $U(1)$ is the only remaining unbroken such selection rule. To arrive back at such a situation one can consider an orbifold of the torus and  insert suitable closed-string twisted mode vertex operators into (\ref{Discchi-main}) which neutralise the individual $U(1)$ charges. Smoothing the orbifold to a generic Calabi-Yau 3-fold corresponds to giving the twisted modes an expectation value, thereby breaking the $U(1)$s and inducing the operator. We describe such neutralising twisted mode insertions in more detail in section \ref{sec:N=1orbvacu}. 

 We conclude that if there are instantons which intersect the required D3 branes, there can be a four-fermion interaction which is allowed by all selection rules and which would lead to a contribution to the superpotential. 
At a very heuristic level, this lifting of the zero-modes $\bar\chi^{\dot\alpha}$ 
can also be understood by viewing the pair $\lambda_{ED} \lambda_{DE}$ as a bosonic operator $X$ with coupling $X (c + \bar\chi \bar\chi)$. Variation with respect to $X$ fixes the E3-position in the sense that $c = - \bar\chi \chi$.
Intuitively, requiring that the E3-brane with bosonic deformation mode passes through the spacetime-filling D3 partially rigidifies the E3-brane. 

Continuity requires that the lifting effect persists even when the D3-brane brane is moved away from the E3-brane. This results in a position dependent mass $m$ for the zero-modes $\lambda_{ED}$ and $\lambda_{DE}$, modifying the effective interaction to become
\be
X (c + m + \bar\chi \bar\chi) \,.
\ee
In this picture the only consequence of $m$ is to shift the relation between $\bar\chi \bar\chi$ and $c$.

Replacing  the modes of type $\bar\chi^{\dot\alpha}$ by the modes $\chi^\alpha$ counted by $H^{2,0}_+(\tilde D)$ in the disk diagram (\ref{Discchi-main}) leads to a correlator whose total worldsheet $U(1)$ charge is non-vanishing because the $U(1)$ charge of  $V^{-1/2}_{\chi}$ is given by  $Q =  - \frac{1}{2}$. 
This prevents a lifting of the $\chi^\alpha$ modes by the same effect as described for the $\bar\chi^{\dot\alpha}$ modes.
This might have been expected already because unlike the $\bar\chi^{\dot\alpha}$, the $\chi^\alpha$ modes are not accompanied by any bosonic deformation modes. Hence there is no analogue of the intuitive picture that requiring the D3-branes to pass through the E3-divisor partially rigidifies the latter.

We have furthermore no reason to believe that the spacetime-filling D3 branes could lift the zero-modes counted by $h^{1,0}_\pm(\tilde D)$ on a smooth Calabi-Yau. Indeed, these are Wilson lines and are therefore protected by a discrete gauge symmetry. 
Note that this conclusion is not in disagreement with the analysis of the toroidal background:
Consider $\mathbb T^2_1 \times \mathbb T^2_2 \times \mathbb T^2_3$ and replace the modes $\bar\chi$ e.g. by the Wilson line modes $\bar\gamma$ associated with the 1-cycles along $\mathbb T^2_2$  (see Table \ref{tab:instmap-main}) in the correlator (\ref{Discchi-main}). This leaves us with $U(1)$ charges $(0,-1,1)$ for the correlator.
This can again be neutralized by insertion of closed string twisted modes on an orbifold. However, for such orbifolds the Wilson line modes are projected out in the first place, in agreement with the fact that the elements of $H^{1,0}(\tilde D)$ of toroidal orbifolds are inherited from the ambient space, while a background preserving  $\cN=1$ supersymmetry has no 1-cycles.

Note that the interaction term (\ref{Sintquartic-main}) is just one of a possible number of four-fermion terms which can lead to instanton contributions to the superpotential. For example, one can have terms involving four fermionic zero-modes of the instanton. However, such a term would require certain conditions on the number of interactions of zero-modes, for example requiring at least two zero-modes.

As alluded to above, we can also interpret the above result from the viewpoint of the F-theory instanton, where rigid divisors on the base contribute to the superpotential.  If there are no rigid divisors, requiring them to pass through points which correspond to some of the positions of the D3 branes in the base of F-theory will rigidify them.  This would be our interpretation of the lifting of the corresponding bosonic zero-modes for deformations of the instanton.  This is similar to what one sees in topological strings and it would be interesting to study the resulting topological theory which captures such instanton contributions in F-theory.  Note also that there were puzzles raised in geometric transitions involving the blow-up of points in F-theory in \cite{Sethi:1996es} and one resolution suggested there was the existence of D3 branes contributing to superpotential terms, in accord to what we have argued above.
%%%%%%%%%%%%%%%%%%%%%%%%%%%%%%%%%%%%%%%%%%%%%%%%
\subsubsection{$\chi(Y_4)=0$ manifolds}
\label{sec_chi0}

The results of the previous section suggest that in searching for clean examples of superpotentials with no instanton corrections we should consider F-theory compactifications on manifolds which have no spacetime-filling D3 branes. This corresponds to the geometric condition of a vanishing Euler number,
\be
\chi(Y_4)=0 \;.
\ee
A list of a class of such Calabi-Yau fourfolds can be found in \cite{Klemm:1996ts}. The examples in this class have $h^{1,1}>20$ which, as discussed in section \ref{sec:instgency}, suggests, at least in a crude first order approximation, that they should exhibit some divisor which can solve (\ref{chitopcond}). An exception to such expectations are Calabi-Yau fourfolds which are a hypersurface in a product of projective spaces, since they were shown to never support an appropriate divisor \cite{Witten:1996bn}. However, all such manifolds with $\chi(Y_4) = 0$ lead to enhanced supersymmetry and do therefore not represent potential counter-examples to our expectations. To see this, we can even consider the more general class of all complete intersection Calabi-Yau fourfolds, which includes the hypersurfaces in products of projective spaces as a special case. These were fully classified in \cite{Gray:2013mja}. It was shown that out of 921,497 such spaces, 15,768 had vanishing Euler number, but they were all product manifolds: ${\mathbb T}^8$, ${\mathbb T}^2\times CY_3$ and ${\mathbb T}^4 \times K3$. These product manifolds actually then lead to enhanced $\cN \geq 2$ supersymmetry.  We will consider other examples of Calabi-Yau fourfolds with $\chi(Y_4) =0$ later in the paper.

Note that another way to avoid D3 branes is by saturating the tadpole through closed string background fluxes. However, turning on background fluxes will induce a superpotential \cite{Gukov:1999ya}. Further, such fluxes can lift instanton zero-modes leading to a non-perturbative contribution to the superpotential (see e.g. \cite{Kallosh:2005gs,Billo:2008sp,Blumenhagen:2009qh,Bianchi:2012kt} and references therein). It is therefore unclear if this can be a viable route towards a vanishing superpotential vacuum.\footnote{If we turn on only D7 worldvolume flux valued in the subspace of $H^{1,1}({\rm D7})$ obtained by pullback from the ambient space, then no superpotential is generated but instead a D-term. However, such flux induces D3 charge on the D7 (as otherwise we could not saturate the D3-brane tadpole with its help), and therefore is naturally expected to have an analogous effect as a D3 brane (see, e.g., \cite{Marchesano:2009rz} in a slightly different context).}

\subsubsection{The $E_8$ superpotential}
\label{sec:e8sup}

The necessity of spacetime-filling D3 branes can also be important in the context of superpotentials which can have instanton contributions but which may still admit supersymmetric Minkowski vacua. An example of such a superpotential was proposed in \cite{Donagi:1996yf} (see \cite{Curio:1997rn} for the heterotic dual). The superpotential was claimed to be an $E_8$ $\theta$-function which was shown to support isolated Minkwoski vacua solutions. The necessary presence of the D3 branes in this example however casts some doubt over whether such solutions exist. This is because including the D3 moduli modifies the superpotential and generally (and possibly inevitably) breaks this $E_8$ structure \cite{Braun:2018fdp}. The fate of the Minkowski vacuum in such a setting remains undetermined, even ignoring fluxed instantons and other effects.

%%%%%%%%%%%%%%%%%%%%%%%%%%%%%%%%%%%%%%%%%%%%%%%%%%%%%%%%%%%%%%%%%
\subsection{Other $\cN=1$ String vacua}
\label{sec:stambvac}

Non-perturbative effects in string theory are not fully understood, and even in concrete examples one may not be able to determine the zero-modes of all instanton branes. This means that there remain classes of string compactifications for which we simply cannot definitively answer the question of whether non-perturbative corrections are present or absent. While in some instances certain corrections can be shown to be absent via dualities, proving that all corrections are absent for these theories will not be possible. In this section we will explain why this is the case, and relegate details to the Appendix. 

% \subsubsection*{Heterotic vacua}

The question of whether there exist four-dimensional $\cN=1$ supersymmetric compactifications of the heterotic string with no worldsheet instanton corrections to the superpotential was actively studied in the early days of the field \cite{Distler:1986wm,Silverstein:1995re,Beasley:2003fx} (see \cite{Buchbinder:2016rmw,Buchbinder:2017azb,Buchbinder:2019hyb,Buchbinder:2019eal} for recent work). The question of superpotential corrections was not revisited with as much intensity after the second superstring revolution despite the fact that now we have a better handle also on spacetime non-perturbative effects. Such spacetime instanton corrections can be due to NS5-branes wrapping the whole Calabi-Yau. Only some of these instantons can be understood as gauge theory instantons belonging to the heterotic gauge group.\footnote{A particularly good handle on this comes from duality with Type I string theory where the instantons are mapped to D5 instantons inside the D9 branes.} This already shows that NS5 instantons may indeed contribute to the superpotential. But it is simple to find examples of heterotic string compactifications which have no gauge instantons, and hence at least these classes of NS5-brane instantons can be absent.

Much more difficult to understand are stringy NS5 instantons, which have no gauge theory interpretation. The central problem is the absence of a microscopic theory of NS5 branes which can be used to calculate the zero-modes. This is further complicated by the fact that the NS5 branes can support bundles of the worldvolume two-form, and the instanton sum must include all such possible bundles. 

To our knowledge, there is no existence proof for stringy NS5 instanton corrections to the superpotential in the heterotic string literature. This leaves open the possibility that there is some obstruction for such corrections to occur. However, in Appendix \ref{sec:stringhet} we present an example of a stringy NS5-brane instanton correction via duality with F-theory. On the F-theory side we consider a Calabi-Yau fourfold constructed as a smooth Weierstrass model over a threefold base given by a $\mathbb{P}^1$ bundle over $\mathbb{P}^2$. Because the fibration is smooth, the heterotic dual has no non-Abelian gauge groups and therefore has no gauge theory instantons. As shown in Figure \ref{fig:instmap}, M5 branes wrapping the base $\mathbb{P}^2$ map to NS5 instantons in the heterotic string \cite{Witten:1996bn}. In Appendix \ref{sec:stringhet} we show that for an appropriate twist of the $\mathbb{P}^1$ bundle, such branes have the correct number of zero-modes to contribute to the superpotential. This serves as a proof of principle that intrinsically stringy NS5 instantons can contribute to the heterotic superpotential, ruling out any potential general obstruction to such contributions. 

The relevance of NS5-brane instantons implies that it is not possible, with current tools, to definitively determine if a given heterotic compactification is free from instanton corrections to the superpotential, at least not directly from the heterotic side. It may be possible to study dual backgrounds in better understood settings, such as F-theory, which brings us back to the discussion of such backgrounds in Section \ref{sec:ftvac}. We will therefore not consider heterotic vacua as possible examples free from corrections. 

This prompts the question whether stringy NS5 instantons are the only effect which precludes an example with no corrections to the superpotential. I.e. if we ignore NS5-brane instantons, are there examples with a vanishing superpotential? In Appendix \ref{sec:hetvac} we study this question, and show that it is very likely that such examples exist. The only ambiguity we could not resolve was due to a new result in \cite{Buchbinder:2019hyb} showing that there exist loopholes in the Beasley-Witten no-go result against worldsheet instanton corrections in certain setups \cite{Beasley:2003fx}. If such loopholes apply to the particular examples we consider in Appendix \ref{sec:hetvac}, then they would rule them out as examples with no corrections to the superpotential except for the question of NS5-brane instantons. In any event, we expect that other examples can be found with no corrections except for stringy NS5-brane effects, and believe that the latter are indeed the only completely general effect which can enforce the presence of a superpotential in generic backgrounds.

In Appendix \ref{App_IIA} we give a brief account of instanton effects in Type I, Type IIA and M-theory compactifications with four-dimensional $\cN=1$ supersymmetry and argue that none of these settings offer sufficient control to definitively prove the absence of instanton corrections to the superpotential.
We also comment on flux compactifications.

\subsection{$\cN =2$ prepotentials and their corrections}
 We now briefly consider the situation in $\cN =2$ theories.
As already discussed in the introduction, we will focus on prepotential terms for $\cN =2$ theories.  In Type IIA theories they have the structure
\be
\cF = \cF_{\mathrm{polynomial}}\left(\Phi\right) + \sum_{n,i} B_{n}\left(\Phi\right) e^{-a^i_n \Phi_i} \;.
\ee
Generically there is no reason for the $B_n$ to vanish.  We ask if this is guaranteed.  As is well-known, general Calabi-Yau manifolds seem to have holomorphic rational curves, and these lead to non-vanishing $B_n$ terms above.  
Indeed  
for a class of manifolds the non-vanishing of $B_n$ is guaranteed on general grounds, in the context of Type IIA compactifications \cite{KamenovaVafa}. While this seems to be the generic feature, the exceptions which arise in the conditions discussed in \cite{KamenovaVafa} do allow possible cases where the $B_n$ may vanish.  Manifolds with vanishing Euler characteristic, $\chi=0$, are part of such an exception.  Indeed we will construct such examples in the next section where we discuss their relation to higher supersymmetric theories.

%%%%%%%%%%%%%%%%%%%%%%%%%%%%%%%%%%%%%%%%%%%%%%%%%%%%%%%%%%%%%%%%%
\section{String compactifications with extra protection}
\label{sec:strnocorr}
%%%%%%%%%%%%%%%%%%%%%%%%%%%%%%%%%%%%%%%%%%%%%%%%%%%%%%%%%%%%%%%%%

The conclusions of section \ref{sec:strpotcor} corroborate our expectation that only in very special $\cN=1$ supersymmetric string compactifications the superpotential can be guaranteed to vanish exactly, and similarly for corrections to the prepotential in $\cN=2$ supersymmetric theories.
In the context of $\cN=1$ theories, we have discussed a number of natural, and previously proposed, candidates for such theories and showed that in all cases some unaccounted effects cast doubt on the conclusion that instanton corrections are really absent. 
One may consider the possibility that theories where the infinite number of potential instanton corrections are absent are in the Swampland. However, this is not the case, and one aim of this section is to prove this by construction. We will discuss various examples where the   $\cN=1$ superpotential is protected against non-perturbative corrections in a way expected for theories with $\cN\geq 2$ supersymmetry.
Similarly we will provide examples of $\cN=2$ theories whose prepotentials receive no corrections in a way one would usually expect for theories with $\cN\geq 4$ supersymmetry.

The basic structure of all the examples we will consider is as follows:  We start with a higher supersymmetric theory and consider orbifolding it by a discrete R-symmetry group $\Gamma$,
which breaks the supersymmetry to the desired one ($\cN=1$ or $\cN=2$).  We find examples where $\Gamma$ has no fixed points and where over all of moduli space the corrections are absent to all orders;  we are able to relate this vanishing to the vanishing properties of the higher supersymmetric theory. 
We also find examples of this type where $\Gamma $ does have fixed points, leading to some additional massless fields.  We show that as long as we do not give a VEV to these massless fields the vanishing condition persists.  But if these fields do acquire a VEV, the protection against the corrections is no longer guaranteed.

Below we give a number of examples of these types of construction for both $\cN=1$ as well as $\cN =2$ theories.   Even though some of the constructions and the arguments for the observed vanishing behaviour are technical (especially in the $\cN=1$ case), they all share the above mentioned simple structure.

%%%%%%%%%%%%%%%%%%%%%%%%%%%%%%%%%%%%%%%%%%%%%%%%%%%%%%%%%%%%%%%%%%%%%%%%%%%%%
\subsection{$\cN=1$ theories}
\label{sec:vann=1}

We begin with theories preserving four-dimensional $\cN=1$ supersymmetry, but which still receive no corrections to the superpotential. There are two types of such constructions, corresponding to whether the superpotential vanishes everywhere in field space or only on a sub-locus of the space of massless fields where it admits a supersymmetric Minkowski vacuum. 
%However, as discussed in the introduction, such a distinction is difficult to formulate sharply for $\cN=1$.

\subsubsection{F-theory vacua with torsional discriminant}
\label{sec:genconst}

Our first example of a class of $\cN=1$ theories with an exactly vanishing superpotential is due to the presence of the Goldstino instanton zero-modes 
 $\bar \tau^{\dot\alpha}$ in Table \ref{tab:instmap-main}. They are related to supersymmetry as follows:

Suppose first that we can take a weakly coupled Type IIB orientifold limit.
As explained in \cite{Blumenhagen:2009qh} and references therein,
the background locally preserves an $\cN=2$ supersymmetry algebra, which can be decomposed schematically as $\cN =2   \supset  (\cN =1)   \oplus  (\cN =1')$.
The system of spacetime filling D-branes and O-planes preserves the first $ \cN =1$ subgroup (with supercharges $Q^\alpha$ and ${\bar Q}^{\dot \alpha}$) and breaks the $ \cN =1'$ superalgebra  (with supercharges $Q'^\alpha$ and ${\bar Q}'^{\dot \alpha}$).
Due to its pointlike localisation within $\mathbb R^{1,3}$, a half-BPS instanton preserves a different ${\cal N}=1$ subalgebra generated by $Q'^\alpha$ and ${\bar Q}^{\dot \alpha}$. The Goldstino modes associated with the spontaneous breaking of the orthogonal complement by the instanton are the modes $\theta^\alpha$ and $\bar \tau^{\dot \alpha}$ as displayed in Table \ref{Goldmodes}.
\begin{table}[t!]
  \centering
  \begin{tabular}{|c|c|}
   \hline   ${\cal N}=1$   &  ${\cal N}=1'$ \\ \hline 
 $\theta^\alpha$    & $\tau^{\alpha}$ \\ \hline
  $\bar\theta^{\dot \alpha}$    & $\bar\tau^{\dot\alpha}$  \\\hline
 \end{tabular}
    \caption{\small Goldstino modes of an instanton in Type II orientifolds.}
   \label{Goldmodes}
\end{table} 

The modes $\bar \tau$ can therefore only be absent if the instanton locally feels the explicit breaking of $\cN =2$ to  $\cN =1$ by the background. This requires, as a necessary condition, that it intersects the system of branes and O-planes suitably.\footnote{There are two possibilities: If the instanton wraps the same cycle as a spacetime-filling brane, the $\bar \tau$ modes are lifted by an ADHM-type interaction \cite{Billo:2002hm,Akerblom:2007uc}, even if it is not a gauge instanton \cite{Aganagic:2007py,Petersson:2007sc}. The second possibility  \cite{Argurio:2007qk,Argurio:2007vqa,Bianchi:2007wy,Ibanez:2007rs} is that the instanton wraps a divisor which, at generic points in its moduli space \cite{Blumenhagen:2007bn,GarciaEtxebarria:2007zv,GarciaEtxebarria:2008pi}, is invariant under the orientifold involution, though not pointwise (a so-called $O(1)$ instanton); this implies that it must intersect the orientifold over a curve.}

There is a more general way to understand when the instanton zero-modes $\bar \tau$ are absent from the F-theory perspective. In F-theory, the system of D-branes and O-planes maps to the discriminant locus $\Delta = 12 \bar K_{B_3}$ (with $\bar K_{B_3} = K^{-1}_{B_3}$ the anti-canonical bundle of $B_3$).
Requiring an intersection of the instanton with the system of branes in F-theory therefore means that (see the discussion in \cite{Cvetic:2009ah,Cvetic:2011gp})
\be
K_{B_3}|_D \neq {\cal O}_D \,.
\ee
If this does not hold, i.e. if $K_{B_3}|_D = {\cal O}_D$, the counting of zero-modes in Table \ref{tab:instmap-main} yields one pair of zero-modes $\bar\tau^{\dot\alpha}$ because ${\rm dim}(H^0(D, K_{B_3}|_D)) =1$.

More generally, such modes are guaranteed to be present for {\it every} instanton divisor $D$ only if $K_{B_3}|_D$ has a section, which means that it must be a non-negative line bundle and hence the anti-canonical bundle must satisfy $c_1(\bar K_{B_3})|_D  \leq 0$, for every divisor class $D$. 
The case $c_1(\bar K_{B_3})|_D  < 0$ for every divisor class $D$ can be excluded because for an elliptic fibration to exist, the anti-canonical bundle $\bar K_{B_3}$ must not be negative.\footnote{Recall that this is because the functions $f$ and $g$ in the Weierstrass model $y^2 = x^3 + f x z^4 + g z^6$ describing an elliptic fibration are sections of $\bar K^4_{B_3}$ and  $\bar K^6_{B_3}$, and therefore $\bar K_{B_3}$ must not be a negative line bundle as otherwise no such sections exist. Likewise, $c_1(\bar K_{B_3})|_D  < 0$ except for some $D$ for which $c_1(\bar K_{B_3})|_D  = 0$ cannot lead to an $\cN=1$ supersymmetric background. }
The only possibility leading to Goldstino modes $\bar\tau$ for every instanton divisor is that $\bar K_{B_3}$ and hence also $K_{B_3}$ is either trivial or torsional. 
Since the first case leads to supersymmetry enhancement in an obvious way, we must require the latter. In this case we must require in addition that $K_{B_3}|_D = {\cal O}_D$ for every divisor $D$ rather than a torsional line bundle on $D$ since such a torsional line bundle has no non-trivial sections.

These two conditions are achieved in F-theory on a {\it freely acting} quotient Calabi-Yau four-fold of the form
\be \label{Y4Z2quotient}
Y_4 = \frac{X_3 \times {\mathbb T}^2}{\Gamma} \,, \qquad \Gamma = \mathbb Z_k \,,
\ee
where $X_3$ is a Calabi-Yau threefold and $k \in\{2,3,4,6\}$. %lead to $\cN=1$ vacua with an exactly vanishing superpotential. 
A simple way to construct such vacua is to let $\Gamma$ act as an involution\footnote{For $\Gamma = \mathbb Z_3, \mathbb Z_4, \mathbb Z_6$ the complex structure of ${\mathbb T}^2$ is fixed at a value compatible with taking the quotient.} on  ${\mathbb T}^2$ and act freely on $X_3$.
  There are many examples of Calabi-Yau threefolds with such freely acting symmetries. The classification of such symmetries for complete intersection Calabi-Yau manifolds reports 166 cases for $\Gamma = \mathbb Z_2$ \cite{Braun:2010vc}. Note that the four-fold $Y_4$ is a non-trivial elliptic fibration over $B_3 = X_3 /\Gamma$.

In fact, for constructions of type (\ref{Y4Z2quotient}), 
\be \label{TorH3B}
{\rm Tor}(H^2(B_3,\mathbb Z)) = \tilde \Gamma \oplus R \,, \qquad \tilde \Gamma = \mathbb Z_k \,,
\ee
where $R$, if non-zero, is inherited from  ${\rm Tor} H^2(X_3,\mathbb Z)$, while $\tilde \Gamma$ is induced by taking the free quotient. Indeed, taking a freely-acting quotient leads to non-trivial torsional one-cycles, and ${\rm Tor}(H_1(B_3,\mathbb Z)) = {\rm Tor}(H^2(B_3,\mathbb Z))$.
Importantly, since $c_1(X_3) = 0$, the only contribution to $c_1(K_{B_3})$ comes from taking the quotient. Therefore
\be
c_1(K_{B_3}) \in \tilde \Gamma \,.
\ee

It follows that the restriction $K_{B_3}|_D$ to any divisor $D$ on $B_3$ must be trivial: Otherwise it would define an element in ${\rm Tor}(H^4(B_3,\mathbb Z))$, but this latter group is unrelated to  taking the quotient by $\Gamma$. Hence, if ${\rm Tor}(H^4(B_3,\mathbb Z))$ is non-trivial, it must be inherited from the parent Calabi-Yau $X_3$ appearing in (\ref{Y4Z2quotient}). This implies that  
$K_{B_3}|_D = {\cal O}_D$.
We have therefore established the presence of the Goldstino zero-modes $\bar\tau$ counted by
\be
{\rm dim}(H^0(D, K_{B_3}|_D)) = 1 \,,
\ee
 preventing the generation of a superpotential everywhere in field space.\footnote{Instead \cite{Blumenhagen:2007bn}, the instantons generate higher derivative F-terms \cite{Beasley:2004ys,Beasley:2005iu} of the form
\be
\int d^4x  d^2\theta  \, \overline{D} \bar{\Phi}  \overline{D} \bar{\Phi} \,  (\ldots) \,  e^{-S(\Psi)} \,.
\ee
Here the $\cN=1$ superfields $\Psi$ and $\Phi$ combine into an $\cN=2$ hypermultiplet in the parent Type II theory prior to orientifolding \cite{GarciaEtxebarria:2008pi}.} 

The same conclusion is reached in the orientifold picture (available for $\Gamma =\mathbb Z_2$) by noting that the double cover of the instanton divisor $D$ 
is the sum of two non-intersecting divisors on $X_3$ exchanged by the 
 freely acting involution. Hence one linear combination of the Goldstino modes survives in F-theory. %  the zero-modes $\bar \tau$ present for the inst

We conclude that instanton corrections to the superpotential are absent everywhere in field space in $\cN=1$ F-theory compactifications when the fibration is purely torsional, i.e. $c_1(K_{B_3}) \in {\rm Tor} H^2(B_3,\mathbb Z)$. 
The reason for the vanishing of the superpotential is not just the presence of arbitrary zero-modes, but of the Goldstino modes associated with an underlying higher supersymmetry.
This is a first encounter for $\cN=1$ theories of the interplay between the absence of non-perturbative effects and the presence of torsional cohomology.  
Note  that the torsional group $\tilde \Gamma$ in (\ref{TorH3B}) translates into a discrete gauge symmetry in the $\cN=1$ F-theory compactification \cite{Camara:2011jg,Mayrhofer:2014laa}. In Section \ref{sec:charexam} we will identify this gauge symmetry as a key common feature at least of the theories with vanishing supersymmetric protected quantities, in this case the superpotential, found in this paper.

\subsubsection{IIB orbifold vacua with $\mathbb{Z}_2$ fixed points}
\label{sec:N=1orbvacu}

{In this section we give an example of an exactly vanishing superpotential on a certain sub-locus of field space.
The absence of the superpotential is due to additional zero-modes for all instantons. As before these can be interpreted as certain Goldstino modes, though this interpretation is less manifest.

The background is chosen as the orientifold of an orbifold ${\mathbb T}^6/\Gamma$. 
The  combined action of $\Gamma$ and the orientifold involution ${\sigma}$ 
breaks the supersymmetry from $\cN=8$ to $\cN=1$. 
This explicit supersymmetry breaking is a global effect, to the extent that locally away from {\it all} fixed point loci of the orbifold and orientifold an object still probes the original supersymmetry.
More generally, if an instanton wraps a cycle which intersects {\it some} of the fixed point loci, it probes only a subset of the original supercharges.
This occurs whenever a subset of the orbifold and orientifold elements acts freely on the instanton. In principle, this subset can be different for different instantons.
As we will propose, the spontaneous breaking of these supercharges by the instanton leads to the appearance of a certain number of instanton Goldstino zero-modes.
These zero-modes are protected except for possible couplings to the massless twisted sector fields of the orbifold. As these acquire a VEV, the zero-modes can be lifted and a superpotential can in principle be generated.

Our notation is as in \cite{Donagi:2008xy}: We consider an orbifold quotient by a discrete group $\Gamma = \bbZ_2 \times \bbZ_2$ acting on the three torus factors of ${\mathbb T}^6 = {\mathbb T}^2_1 \times {\mathbb T}^2_2 \times {\mathbb T}^2_3$ with complex coordinates $\left(z_1,z_2,z_3\right)$.  The action of a $\bbZ_2$ is specified by $\left(\epsilon_1 \delta_1,\epsilon_2 \delta_2,\epsilon_3 \delta_3 \right)$ with  $\epsilon_i \in \{0,\frac12\}$ 
and $\delta_i = \pm$ as
\be
z_i \rightarrow \delta_i \, z_i + \epsilon_i \,.
\ee
% denoting the shift on the coordinates (with periodicity 1) and $\delta_i = \pm$ the sign action of the coordinate. 
The complex coordinates decompose into real components as $dz_i = dx_i + \tau dx_{i+3}$ (we take $\tau=\sqrt{-1}$ unless otherwise specified).

We consider a Type IIB orientifold of the orbifold modded out by the $\bbZ_2$ symmetries generated by
\be 
\theta \;: \; \left(0+,0-,0-\right) \;, \qquad \theta' \;: \; \left(0-,0+,0-\right) \;\; \;.
\label{iiborb}
\ee
Unlike \cite{Vafa:1994rv}, where this orbifold was studied, we do not turn on discrete torsion, which means the Hodge numbers are $h^{1,1}=3$ and $h^{2,1}=51$. These arise from $h^{1,1}=h^{2,1}=3$ universal modes of $\mathbb T_1^2 \times \mathbb T_2^2 \times \mathbb T_3^2$ and 48 twisted modes. We then further orientifold by a holomorphic involution (see for example \cite{Angelantonj:1999ms,Blumenhagen:2003vr})
\be
{\sigma} \;: \; \left(0-,0-,0-\right) \;.
\ee
This leads to 64 O3-planes located at the fixed points of ${\sigma}$, as well as 12 O7-planes located at the fixed points of ${\sigma} \theta$, ${\sigma} \theta'$ and ${\sigma} \theta\theta'$. We place 8 D7 branes on top of each of the O7 planes to cancel the tadpoles, giving a gauge group $SO(8)^{12}$. We also need to add 32 D3 branes to cancel the O3 charges, which can be placed anywhere on the orbifold.

The possible instantons contributing to the superpotential come from D3 branes wrapping holomorphic divisors. Since $h^{1,1}=3$, these instantons are associated purely to the untwisted sector. 
The generators of the divisor group are the four-cycles on ${\mathbb T}^6$ which are not projected out by the quotient by $\Gamma$ and ${\sigma}$.

To study these, let us introduce the notation
\be
D_i = \frac{1}{|\Gamma|} \sum_{g \in \Gamma} g \, {\mathbb D}_i \,, \qquad \quad {\mathbb D}_i = p_i \times  {\mathbb T}^2_{j} \times {\mathbb T}^2_{k}     \,,
\ee
where $\mathbb D_i$  is a divisor on ${\mathbb T}^2_1 \times {\mathbb T}^2_2 \times {\mathbb T}^2_3$ located at the point $p_i$ on ${\mathbb T}^2_i$ (and $(i,j,k)$ is a permutation of $(1,2,3)$).
Then the divisor group of the orientifolded background is generated by
\be \label{calD-Def}
{\cal D}_i = \frac{1}{2} (D_i +  D_i') \,, \qquad \quad D_i' = {\sigma} \, D_i \,.
\ee
Note in particular that the orientifold image of the point $p_i$ is the reflected point $p_i' = {\sigma} p_i$, and $p_i' \neq p_i$ unless $D_i$ is located at one of the 4 orientifold fixed planes on ${\mathbb T}^2_i$.

Before discussing more general cycles than the generators ${\cal D}_i$, let us note that
for generic position of $p_i$, a subset of the orbifold and orientifold group elements acts freely on ${\cal D}_i$.
For example, consider the divisor ${\cal D}_1$, wrapping the second and third torus with $p_1$ away from the fixed points on ${\mathbb T}^2_1$. The freely acting elements are those which 
act as inversion on ${\mathbb T}^2_1$ and hence map $p_1$ to $-p_1$.
These are the elements
$\theta'$, $\theta \theta'$, ${\sigma}$ and ${\sigma} \theta$.
Note that these include the generators of the orientifold involution, $\sigma$, and of one of the $\mathbb Z_2$ orbifold factors, $\theta'$, even though the set of freely-acting elements on the instanton does not form a group
 $\mathbb Z_2 \times \mathbb Z_2$.
By contrast, the non-freely acting group elements are those which act as inversion along
${\mathbb T}^2_2 \times {\mathbb T}^2_3$ or separately on ${\mathbb T}^2_2$ and ${\mathbb T}^2_3$ without affecting ${\mathbb T}^2_1$. These are the elements $\theta$, ${\sigma} \theta'$ and $({\sigma} \theta')\,  \theta$.

The fact that some elements of the combined orbifold and orientifold action act freely on the instanton implies that the instanton locally probes a higher supersymmetry than the 
$\cN=1$ supersymmetry. This should result in the appearance of Goldstino instanton zero-modes beyond the universal modes $\theta^\alpha$ because the instanton spontaneously breaks this higher supersymmetry.
The situation is similar to the appearance of Goldstino modes in the setup of Section \ref{sec:genconst}, with the difference that first we expect more Goldstino modes since the instanton probes a higher local supersymmetry than in this example, and second the specific embedding of the higher supersymmetry into the original $\cN=8$ differs for the different choices of cycles ${\cal D}_i$, $i=1,2,3$.

In fact, the instanton zero-modes for an instanton along ${\cal D}_i$ at generic position are the modes
\bea \label{DiDi'modes}
(x^\mu, \theta^\alpha, \bar\tau^{\dot\alpha}) \,, \qquad 
(c_i, \chi_i^\alpha, \bar\chi_i^{\dot\alpha}) \,.
\eea
Apart from the four universal fermionic modes $\theta^\alpha$ and $\bar\tau^{\dot \alpha}$ the instanton exhibits four fermionic deformation modes associated with its location on ${\mathbb T}^2_i$. Note that the zero-modes associated with the Wilson lines along the wrapped tori prior to taking the quotient are projected out.

A general D3-brane instanton wraps a divisor of the form
\be
{\cal D} = \sum_{i=1}^3 a_i {\cal D}_i \,.
\ee
If each $D_i$ and $D_i'$ in (\ref{calD-Def}) are locally identical, ${\cal D}$ is invariant pointwise under the orientifold (corresponding to an {\it Sp(1) instanton}).\footnote{The contribution of such an instanton to the superpotential, if non-vanishing, involves the $SO(8)$ gauge sector fields located at the orientifold planes. Apart from this, the relevant aspects of the following discussion about the lifting of the remaining zero-modes applies to this type of instantons as well and we therefore do not need to consider these separately.}
 % , while away from the fixed points, 
For more general ${\cal D}$, the 
situation is similar to the individual ${\cal D}_i$ at generic position discussed above, and the instanton in particular always exhibits the zero-modes $\bar\tau^{\dot\alpha}$ which must lifted for a superpotential to be generated. 
% ${\cal D}$ is mapped to its orientifold image ${\cal D}' = \sum_{i=1}^3 a_i D_i'$. Here $D_i' = p'_i \times  {\mathbb T}^2_{j} \times {\mathbb T}^2_{k}$, where $p_i' = {\sigma} \, p_i$  is the image of $p_i$ under the orientifold involution on ${\mathbb T}^2_i$.
To see when this can happen, note that if at least two $a_i$ are non-zero, one can deform the formal sum of divisors in ${\cal D}$
into single divisor which is invariant as a whole (though not pointwise) under the orientifold involution. 
For the potential contribution to the superpotential it is immaterial in which phase we consider the instanton (in fact, integrating over the instanton moduli space requires us to sum up both types of contributions).
We therefore focus on the generic situation, corresponding to a smooth invariant divisor, which we call $\tilde {\cal D}$.

The invariant bound state $\tilde {\cal D}$ locally feels the orientifold action and we are left with only one universal mode system $(x^\mu, \theta^\alpha)$ as well as a certain number of orientifold odd components of type $(c, \bar\chi^{\dot\alpha})$. The latter are, in this sense, remnants of Goldstino zero-modes of the constituent cycles $D_i$ and $D_i'$. This implies that we expect them to be protected except for possible couplings involving the massless twisted closed string fields.
Once the latter acquire a VEV, the local structure of supersymmetry is broken and the zero-modes can in principle be lifted.\footnote{Note that the closed string sector massless modes correspond to complex structure deformations of the orbifold.}
 Therefore the interactions responsible for the lifting of the Goldstino modes must involve the twisted massless fields, and the superpotential must have 
 twisted-mode prefactors. 
Since we must introduce spacetime filling D3 branes to cancel the O3 charges, the deformation modes could, for instance, be lifted by the mechanism of section \ref{sec_P1P2}. But, again, this would lead to twisted modes as instanton prefactors.

The crucial question is how many twisted mode expectation values are required to lift all the zero-modes. Suppose that all instantons could contribute to the superpotential with only two or more twisted mode insertions; under this assumption there would exist a supersymmetric Minkwoski vacuum with an exactly vanishing superpotential. Specifically, all instanton terms in the superpotential would take the form
\be
\left( \prod_{p} \Phi_p^{i_p} \right) e^{-T} \;,
\label{iibsup}
\ee
where $T$ denotes the superfield controlling the volume of the cycle wrapped by the instanton. The $\Phi_p$ are twisted modes, which must be inserted $i_p$ times, and the worldsheet $U(1)$ charge selection rules discussed in section \ref{sec_P1P2} would imply that, by the above assumption 
\be
\sum_{p} i_p \geq 2 \;.
\label{2ormoretm}
\ee
The superpotential (\ref{iibsup}) admits a solution to $W=dW=0$ on the locus $\Phi_p=0$. Note that the existence of this vacuum is independent of the K\"ahler potential, which means that we need not worry about how to calculate the kinetic terms for the twisted modes. The rest of this section is dedicated to showing that the condition (\ref{2ormoretm}) indeed holds.

We will only be concerned with the charges of the vertex operators under the three worldsheet $U(1)$ selection rules associated to the tori. 
In the $(-1/2)$ picture, the vertex operators e.g. for deformation modes along ${\mathbb T}^2_1$ %or Wilson line modes along ${\mathbb T}^2_2$ 
have the following charges (see e.g. \cite{Cvetic:2007ku}):
\be
V_{\bar\chi_1}^{(-1/2)}\;:\; \left(-\frac{1}{2}, \frac12,\frac12 \right) \,.%\,, \qquad V_{\bar\gamma_2}^{(-1/2)}\;:\; \left(\frac{1}{2}, -\frac12,\frac12 \right)  \,.
\ee
The massless closed-string twisted modes in the $(-1,-1)$ picture have charges (see, for example \cite{Conlon:2011jq})
\be
V^{(-1,-1)} \;:\; \left(2\theta_1,2\theta_2,2\theta_3\right) \;,
\ee
where the $\theta_i$ are the orbifold twists and $\sum_i \theta_i = 1$. The overall factor of 2 appears because for closed-string operators the twist fields appear in both left and right-moving sectors. So in the case of the orbifold (\ref{iiborb}) we have in particular
\be
V_1^{(-1,-1)} \;:\; \left(1,1,0\right) \;, \qquad V_2^{(-1,-1)} \;:\; \left(1,0,1\right)  \;, \qquad V_3^{(-1,-1)} \;:\; \left(0,1,1\right)  \;.
\ee
To achieve charge $-2$, as required for disc diagrams,  we may have to picture change some of the vertex operators. This is done by the prescription \cite{FMS} (see also for example \cite{Conlon:2010qy,Conlon:2011jq})
\be
V^{(c+1)}\left(w\right) = \lim_{z \rightarrow w} e^{\phi(z)} T_F(z) V^{(c)}(w) \;,
\ee
where $\phi$ is the (bosonised) ghost, and the (internal part of) $T_F$ is
\be
T_F(z) = \sum_{i=1}^3 \left[ \partial \bar{z}_i \psi_i(z) + \partial z_i \bar{\psi}_i \right] \;.
\ee
Here the $\psi_i$ and $\bar{\psi}_i$ are the bosonised worldsheet spinors along the torus directions and have charges $+1$ and $-1$, respectively. At least purely with respect to the $U(1)$ charges, each picture changing adds $\pm1$ to one of the $U(1)$ charges.

We would like to consider lifting the deformation modes $\bar{\chi}_1$ along ${\mathbb T}^2_1$ through some operator
\be
 \cO \bar{\chi}_1 \bar{\chi}_1 \;.
 \label{liftcb}
 \ee
The piece $\bar{\chi}_1\bar{\chi}_1$ has $U(1)$ charges $\left(-1,1,1\right)$ and ghost charge $-1$. It is therefore not possible for $\cO$ to be an insertion of just a single twisted vertex operator in the appropriate $(0,-1)$ picture. Further, since each instanton has just one deformation mode per torus, it is not possible to write a quartic interaction. 
On the other hand,  $\cO$ can in principle correspond e.g. to a term quadratic in the massless closed string twisted vertex operators, at least to the extent that such correlators do not violate any worldsheet $U(1)$ charge.

In addition to such direct couplings to the closed string massless twisted sector, 
couplings involving the spacetime filling D3 branes could lift the deformation modes $\bar\chi_i$. Indeed, if we consider the $(0,0)$ picture twisted vertex operators
\be
V_1^{(0,0)} \;:\; \left(1,0,-1\right) \;, \qquad V_2^{(0,0)} \;:\; \left(1,-1,0\right) \;,
\ee
then the disc correlator
\be
\left<V_1^{(0,0)}  V_2^{(0,0)}  V^{-1/2}_{\lambda_{DE}}V^{-1/2}_{\bar\chi}V^{-1/2}_{\bar\chi}V^{-1/2}_{\lambda_{ED}}\right>_{Disc} 
\label{disctwis}
\ee
is neutral under all the $U(1)$ charges. This gives an explicit realisation of the effect discussed in section \ref{sec_P1P2} (but does not ensure the further required condition of a non-trivial zero momentum limit). Again we see that the operator $\cO$ in (\ref{liftcb}) involves at least two twisted mode insertions, thereby ensuring  (\ref{2ormoretm}).

We conclude that all instanton contributions to the superpotential are such that $W = dW = 0$ on the sub-locus in moduli space where the (closed string) massless twisted sector fields take a vanishing expectation value. 
The origin of this vanishing result are the zero-modes which are related to the individual deformations of the divisors ${\cal D}_i$ generating the divisor group of the orbifold.
We interpret these zero-modes as Goldstino modes of the instanton spontaneously breaking a certain higher supersymmetry probed locally by the instanton.

{
It is instructive to contrast the orbifold action  (\ref{iiborb}) without discrete torsion to its cousin with discrete torsion \cite{Vafa:1994rv}, giving rise to Hodge numbers 
$h^{1,1} = 51$, $h^{2,1} = 3$. 
The twisted sector now includes rigid divisors without any deformation  moduli. Instantons along such divisors contribute to the superpotential \cite{Denef:2005mm,Lust:2005dy} even without the need to invoke the mechanisms of zero mode saturation described above. See \cite{Blumenhagen:2005tn,Cvetic:2007ku} for the mirror dual Type IIA setting, and \cite{Lust:2005dy,Lust:2006zg} for a systematic study of other Type IIB orbifolds with similar properties. 
The resulting superpotential depends {\it exponentially} on the twisted sector {K\"ahler} (as opposed to complex structure) moduli. It does therefore {\it not} vanish on the sub-locus in moduli space where the latter fields take a zero value. 
All this is in agreement with our interpretation of the deformation zero-modes as Goldstino modes in the model without discrete torsion: Since the rigid divisors in presence of discrete torsion are stuck at the orbifold fixed loci, they do not probe a local higher symmetry except possibly the one preserved away from the orientifold fixed planes. There is therefore no rationale for them to exhibit Goldstino zero-modes other than the modes $\theta^\alpha$ and $\bar\tau^{\dot\alpha}$; the latter are absent as long as the instanton divisor intersects one of the orientifold fixed planes in a suitable manner, as before.  
}

%%%%%%%%%%%%%%%%%%%%%%%%%%%%%%%%%%%%%%%%%%%%%%%%%%%%%%%%%%%%%%%%%%%%%%%%%%%%%
\subsection{$\cN=2$ theories}
\label{sec:vann=2}

For $\cN=2$ supersymmetry, the supersymmetric protected quantity  is the prepotential and we are interested in theories where (\ref{bn0}) holds. In this setting we can also sharply differentiate between the condition (\ref{bn0}) holding over all the vector multiplet moduli space, or only on a sub-locus. %We first present an example of the latter, and then move on to the former. 

\subsubsection{Type II string theory on orbifolds}
\label{sec:iiaorbN2}

Consider Type II string theory on ${\mathbb T}^6$ or $K3\times {\mathbb T}^2$, leading to $\cN=8$ and $\cN=4$ supersymmetry, respectively.
Type II string theory on orbifolds of these spaces is an example where the prepotential is purely polynomial, in suitable coordinates, at least 
 on a sub-locus of moduli space.   To begin with, it is well known that the prepotential for $\cN=8,4$ supersymmetric theories is purely cubic.  However we now argue this is still true at least for some sub-loci of any orbifold of these spaces which breaks supersymmetry to $\cN=2$.
 
 To see this, note that the prepotential, which captures the low energy sector of vector multiplets in Type II string theory, is computed at string tree-level, i.e. at genus-zero on the string worldsheet.  This is because vector multiplets and hypermultiplets decouple in $\cN=2$ theories, and for Type II theories the coupling constant is given by the expectation value of a hypermultiplet and so we can take the weak coupling limit and compute the prepotential exactly by computing at string tree-level. Since the genus-zero amplitudes of the orbifold theory restricted to the untwisted vertex operators are identical to that of the unorbifolded theory (as there are no non-trivial cycles to include orbifold holonmy twists), we see that restricting the prepotential to the untwisted fields gives only a cubic polynomial as in the higher supersymmetric theory.   If there are no massless fields in the twisted sectors this proves that the prepotential is exactly cubic for all moduli fields, because those all come from the untwisted sector.  However, if there are massless fields in the twisted sectors, it is no longer true that including those moduli in the prepotential will lead to a cubic prepotential: This is because we now need to compute amplitudes involving vertex operators of twisted fields, which receive no protection from a higher supersymmetric theory, to which they do not belong.
This means that in such a case, only if we set the
 expectation value of massless twisted states to zero are the amplitudes unmodified. Hence the prepotential is polynomial  only on the sub-locus of vanishing twisted sector moduli.
 
 Note that this argument critically uses the fact that the prepotential is generated at string tree-level in Type II string theory.
For higher genus amplitudes, computing $F^{(g)}$ with $g>0$, it is no longer true that the relevant worldsheet has no one-cycles and so even the amplitudes involving untwisted vertex operators are no longer protected.  An example of this is the correction to genus-one topological amplitudes discussed in \cite{Bershadsky:1993cx}.  As shown there, these amplitudes vanish for theories with ${\cN =8}$ supersymmetry exemplified by Type II compactifications on ${\mathbb T}^6$.  However they do not vanish on  ${{\mathbb T}^4}/{\mathbb{Z}_2} \times {\mathbb T}^2=K3\times {\mathbb T}^2$ or ${\mathbb T}^6/\Gamma$ (for generic $\Gamma\subset SU(3)$), which have ${\cN =4}$ or ${\cN =2}$ supersymmetry, even on the locus where twisted fields are set to zero.  

Similarly, the argument does not imply the vanishing of the prepotential corrections e.g. for the heterotic string on a freely acting orbifold ${\mathbb T}^6/\Gamma$ because  the prepotential receives corrections at one-loop level in the heterotic frame.
 Indeed,  suppose we take a freely acting $\mathbb Z_2$ orbifold of ${\mathbb T}^4 \times {\mathbb T}^2$ with the standard embedding, leading to gauge group $[E_7 \times SU(2)] \times E_8$. At the massless level, the orbifold projects out, in particular, part of the states in the adjoint representation of the $E_8$ gauge factor. This leads to non-vanishing corrections to the prepotential at one-loop, in agreement with the above argument. We conclude that the appearance of an orbifold structure for an arbitrary string theory as such is not sufficient to guarantee the vanishing of the prepotential, be it everywhere in moduli space or only on a sub-locus.  The same applies to ${\cN =1}$ amplitudes.
 Thus the orbifold structure we have found is sufficient to guarantee protection against corrections only in specific cases, and is not sufficient for protection in a general situation.

\subsubsection{IIA string theory on the Enriques Calabi-Yau}
\label{sec:enrcy}

For further illustration, and also to make contact with the discussion in Section \ref{sec:vann=1}, note that Type IIA string theory on the Enriques Calabi-Yau is an example of an $\cN=2$ theory where the property (\ref{bn0}) is satisfied over all the moduli space.\footnote{The first study of this Calabi-Yau was in \cite{s}; in particular all genus-zero Gromov-Witten invariants vanish \cite{Klemm:2005pd}, which means that there are no worldsheet instanton corrections to the Type IIA prepotential. A closely related study of Type II string theory on this manifold has also appeared in \cite{KashaniPoor:2013en}.} 
The Calabi-Yau $X_3$ is the smooth resolution of the orbifold 
\be
\theta \;: \; \left(0+,0-,0-\right) \;, \qquad \theta' \;: \; \left(0-,\frac12 +, \frac12 -\right) \;\; \;.
\label{enorb}
\ee
It is also possible to smooth out the first $\mathbb{Z}_2$ and write $X_3$ as a freely acting orbifold 
\be \label{X3as quotient}
X_3 = \frac{{\mathbb T}^2 \times K3}{\mathbb{Z}_2} \;.
\ee
Because the orbifolding is freely acting, the Enriques Calabi-Yau has holonomy $SU(2)\times \mathbb{Z}_2$ rather than $SU(3)$. 
By the general argument of Section \ref{sec:iiaorbN2}, the worldsheet instanton corrections to the prepotential must be vanishing, and in fact in the entire moduli space because there exists no massless twisted sector since the involution is freely-acting.

{Alternatively, we can understand the absence of prepotential corrections in a manner similar to the discussion in Section \ref{sec:genconst} by noting that every worldsheet instanton on $X_3$ necessarily has too many zero-modes to contribute to the prepotential.
As in the $\cN=1$ context, the presence of these universal zero-modes is related to supersymmetry. Technically it is a consequence of the fact that the Enriques Calabi-Yau is elliptically fibered over the Enriques surface $B_2$ with
\be\label{KB2Enriques}
c_1(K_{B_2}) \in {\rm Tor} H^2(B_2,\mathbb Z) = \tilde \Gamma = \mathbb Z_2 \,.
\ee
Note that this relation implies that also ${\rm Tor} H^2(X_3,\mathbb Z) = \mathbb Z_2$. 
Duality between Type IIA on $X_3$ with base $B_2$ and F-theory on $B_2 \times {\mathbb T}^2$
maps a worldsheet instanton on a curve $C$ on $B_2$ to a
 D3-brane along $C \times {\mathbb T}^2$.
A D3-brane wrapping the curve $C$ in F-theory on $B_2$ gives rise to a string in six dimensions. Its worldsheet theory contains a collection of $4 h^0(C, K_{B_2}|_C) = 4$ universal zero-modes as well as $4 h^0(C, K_{B_2}|_C)$ extra modes \cite{Haghighat:2015ega,Lawrie:2016axq}, which we interpret as Goldstino modes.\footnote{The first type of modes are the ones called $\mu_+$ and $\tilde \mu_+$ in Table 3 of \cite{Lawrie:2016axq} in representation $({\bf 2},1)$ of $\mathbb R_{T}^4$ and the latter are the modes $\lambda_-$ and $\tilde \lambda_-$ in representation $(1,{\bf 2})$. Here $\mathbb R^4_T$ refers to the four directions transverse to the string in $\mathbb R^{1,5}$.} 
In reducing the string along ${\mathbb T}^2$ we obtain a worldsheet instanton in Type IIA string theory with corresponding zero-modes. These are the $\cN=2$ analogue of the universal modes $\theta^\alpha$ and, respectively, the modes $\bar\tau^{\dot \alpha}$ in the ${\cN}=1$ context.
In particular there exist four additional zero-modes, analogous to the modes $\bar\tau$, whenever  $K_{B_2}|_C = {\cal O}_C$. For $X_3$ of the form (\ref{X3as quotient}) this is guaranteed for every curve $C \in B_2$ for the same reasons as in Section \ref{sec:genconst}. 
 }

%%%%%%%%%%%%%%%%%%%%%%%%%%%%%%%%%%%%%%%%%%%%%%%%%%%%%%%%%%%%%%%%%%%%%%%%%%%%%
% \subsection{$\cN=4$}
% \label{sec:vann=4}

%%%%%%%%%%%%%%%%%%%%%%%%%%%%%%%%%%%%%%%%%%%%%%%%%%%%%%%%%%%%%%%%%%%%%%%%%%%%%
\section{Characteristics of the examples with extra protection}
\label{sec:charexam}

We have shown that there are a number of examples, and quite general constructions, which do not receive instanton corrections to their supersymmetric protected quantities, but preserve less than the expected amount of supersymmetry for such protection. It is natural to consider the general features of such examples, and how these may be responsible for this surprising amount of protection. In this section we will identify such general features. % and propose a swampland conjecture which follows from them. 

Let us first introduce some notation. If in a supergravity theory a supersymmetric protected quantity, in the sense of the Introduction, vanishes, we say that the theory  {\it mimics} a higher supersymmetric theory. For instance, if a theory with $\cN=1$ supersymmetry has a vacuum with an exactly vanishing potential, it mimics an $\cN=2$ theory.

Our claim is that all the examples constructed in this work, where a theory mimics one with higher supersymmetry, take the following form: There are two supergravity theories, %Theory $A$ and Theory $B$, preserving the following amount of supersymmetry:
\be
\begin{aligned}
&{\rm Theory \, \, A}: \qquad \quad \cN=k'        &{\rm supersymmetric} \,, \cr 
&{\rm Theory \, \, B}: \qquad \quad \cN=k < k' &{\rm supersymmetric}  \,.
\end{aligned}
\ee
Theory $A$ has an R-symmetry group  containing a discrete subgroup $\Gamma$, which is necessarily gauged because there are no global symmetries in quantum gravity (in more than two spacetime dimensions).
Theory $B$ is obtained as the orbifold quotient 
\be
B = A/\Gamma \,.
\ee
As always in the context of (abelian) orbifolds (see for example \cite{Ginsparg:1988ui}), the orbifold Theory $B$ exhibits a discrete gauge symmetry $\tilde \Gamma$ such that the neutral sector of $B$ with respect to $\tilde\Gamma$ coincides with the neutral sector of $A$ under $\Gamma$.\footnote{If $\Gamma$ is abelian, as in all examples considered in this paper, then $\Gamma$ and $\tilde \Gamma$ are isomorphic as groups.} 
Note, however, that $\tilde \Gamma$ is not an R-symmetry group of $B$. The neutral sector of Theory $B$ under symmetry $\tilde \Gamma$ is hence embeddable into Theory A, as the sector neutral under $\Gamma$.
On the locus in the field space of $B$ where the symmetry $\tilde \Gamma$ is unbroken, the theory {\it mimics} aspects of a higher supersymmetric theory (more precisely the structure of an $\cN=k'$ supersymmetry as in Theory $A$).

% There is a discrete element of the higher supersymmetry R-symmetry group $\mathbb{Z}_p$. In the lower supersymmetry theory $B$, this discrete symmetry is gauged (but is not a subgroup of the lower supersymmetry R-symmetry). The neutral sector of the $B$ theory under this symmetry is embeddable in the higher supersymmetric theory $A$. 

The cases where Theory $B$ mimics the structure of higher supersymmetry over all of its moduli space, for example when the prepotential is cubic over all the vector multiplet moduli space, is a further restriction of this scenario. In such cases all the massless fields in Theory $B$ are neutral under $\tilde \Gamma$.

This structure is indeed realized in all the examples of theories mimicking higher supersymmetry in this paper.
The sector of Theory $B$ charged under $\tilde \Gamma$ is identified, in orbifold language, with the twisted sector labeled by group elements of $\Gamma$.
In the presence of such a twisted sector, e.g. for $\Gamma= \mathbb Z_2$, Theory $B$ has an exact $\tilde \Gamma =\mathbb{Z}_2$ symmetry which acts by sending all the twisted modes to minus themselves.   Modding out the orbifolded theory by this gauged $\tilde \Gamma = \mathbb{Z}_2$ gives back the original theory.  The sector of the orbifold theory invariant under $\tilde \Gamma$ comes from the untwisted sector of theory $A$, which was invariant under $\Gamma$.

The symmetry group $\tilde \Gamma$ expected in orbifold constructions can also be understood from geometric considerations in the cases without a {\it massless} twisted sector, where we have a freely acting quotient operating on Theory $A$ by a discrete symmetry $\Gamma$. Taking the quotient leads to a compactification manifold $M$ with non-trivial 
torsional group\footnote{If the compactification underlying Theory $A$ has torsion itself, the freely acting quotient ${\it adds}$ the torsion factor $ \Gamma$ to this independent torsional group.} ${\rm Tor}H_1(M,\mathbb Z) = {\Gamma}$.  Non-trivial elements of this ${\rm Tor}H_1$ precisely label the twisted sectors of the orbifold, i.e. the winding string states on these 1-cycles.  $\tilde \Gamma$ is dual to this and can be viewed as the space of representations of $\Gamma$.
 Such torsional cohomology implies a discrete gauge symmetries in  Theory $B$ \cite{Camara:2011jg,Mayrhofer:2014laa} which in this case is nothing but the usual symmetry group associated to the existence of twisted sectors of the orbifold $M$ and their symmetry under interactions given by $\tilde \Gamma$ (see e.g. \cite{Hamidi:1986vh}).

%%%%%%%%%%%%%%%%%%%%%%%%%%%%%%%%%%%%%%%%%%%%%%%%%%%%%%%%%%%%%%%%%%%%%%%%%%%%%
\section{Discussion}
\label{sec:dis}
%%%%%%%%%%%%%%%%%%%%%%%%%%%%%%%%%%%%%%%%%%%%%%%%%%%%%%%%%%%%%%%%%%%%%%%%%%%%%

In this work we have argued that the superpotential and prepotential in a quantum gravity theory with four-dimensional $\cN=1$ and $\cN=2$ supersymmetry receive non-perturbative corrrections unless they are protected by supersymmetry, in the sense that the theory is related in a subtle way to a higher supersymmetric theory.

{In all examples where we have observed this phenomenon, the relation to the higher supersymmetric theory is essentially via an orbifolding procedure. In absence of a twisted massless sector, the protection 
by this higher supersymmetry
is at work all over moduli space, while more generally it can be broken if the massless twisted sector fields acquire a vacuum expectation value. 
In all examples, we have given various arguments for this behaviour. In the context of Type II compactifications with $\cN=2$ supersymmetry, the absence of corrections to the prepotential could be explained by general properties of genus-zero correlators in orbifolded theories.
In other string theoretic examples, in particular in models with $\cN=1$ supersymmetry,
 we have traced back the vanishing of the superpotential behaviour to the appearance of certain instanton zero-modes: These are to be interpreted as Goldstino modes associated with the spontaneous breaking of a higher supersymmetry probed locally by the instantons.
The explicit breaking of the higher supersymmetry by the orbifolding is global in nature. As long as an instanton does not intersect any of the fixed loci of the orbifold, it probes the higher supersymmetry and therefore exhibits certain Goldstino modes in its worldvolume. The latter are lifted only via interactions involving the massless twisted sector fields, if present.
}

{
While in all the examples with a non-generic vanishing behaviour which we have found we have observed an orbifold structure,
the latter does not in general
 guarantee a protection, for instance, of the superpotential or prepotential against corrections. 
We have exemplified this by recalling that e.g. the ${\mathbb T}^6/\mathbb Z_2 \times \mathbb Z_2$ orbifold with discrete torsion \cite{Vafa:1994rv} has a non-zero superpotential \cite{Denef:2005mm,Lust:2005dy,Lust:2006zg}
generated by instantons along blow-up divisors throughout its moduli space.
This is in full agreement with our interpretation of the unliftable instanton zero-modes as Goldstino modes. Similarly in the $\cN =2$ case, the heterotic string on a freely-acting orbifold can exhibit one-loop corrections to the prepotential, unlike our protected examples in the context of Type II theory, which are exact at genus zero.} 

Our results for theories with $\cN=1$ and $\cN=2$ theories in four dimensions prompt the question whether there can also be quantities in an $\cN=0$ supersymmetric theory, such as the potential, which vanish if the theory is related to an $\cN=1$ theory in a similar manner.
Another candidate for a partly protected quantity arises in the context of the Weak Gravity Conjecture \cite{ArkaniHamed:2006dz}, which states that every gauge theory coupled to quantum gravity must exhibit some super-extremal particle of charge $(q,m)$ satisfying
\be
\frac{q^2}{m^2} \geq  \frac{c}{M^{d-2}_{\rm Pl}} \,.
\ee
Here the order one constant $c$ depends on the number of spacetime dimensions $d$ and the details of the theory such as presence of massless scalar fields \cite{Heidenreich:2015nta,Palti:2017elp,Lee:2018spm}.
According to a strong form of this conjecture \cite{Ooguri:2016pdq}, the equality should hold for a particle exactly including quantum corrections only if the theory is supersymmetric and the state is BPS: Otherwise there is no protection against slight quantum corrections taking the particle away from the strict equality.   In a sense this is similar to the general lesson of this paper that with lower supersymmetry everything that is allowed will happen.  Taking this analogy further, even though we are not aware of any counterexamples to this conjecture, in the spirit of this paper one might wonder if there can exist exceptional non-supersymmetric theories in which nonetheless strict equality for the non-BPS WGC states holds, possibly again because of a subtle relation to a higher supersymmetric theory.

\noindent
{\bf Comment:} While this work was in completion we were notified of the work \cite{cecotti}, which has some overlap with our ideas.
\newline
\newline
\noindent
{\bf Acknowledgements:} We would like to thank Ralph Blumenhagen, Sergio Cecotti, Albrecht Klemm, Dieter L\"ust and Stephan Stieberger for helpful discussions and correspondence.  We have greatly benefited from the hospitality of UC Santa Barbara KITP where this project was completed.

 The research of C.V. is supported in part
by the NSF grant PHY-1719924 and by a grant from the
Simons Foundation (602883, CV).   This research was supported in part by the National Science Foundation under Grant No. NSF PHY-1748958.

%%%%%%%%%%%%%%%%%%%%%%%%%%%%%%%%%%%%%%%%%%%%%%%
\appendix

%%%%%%%%%%%%%%%%%%%%%%%%%%%%%%%%%%%%%%%%%%%%%%%%%%%%%%%%%%%%%%%%%%%%%%%%%%%%%
\section{Flux lifting of D3-instanton zero-modes in F-theory} \label{appendix_Flux}

In this appendix we discuss the lifting of D3-brane instanton zero-modes in F-theory by instanton flux. After briefly reviewing the general mechanism of \cite{Bianchi:2011qh} in Appendix \ref{subsec_genflux}, we argue in Appendix \ref{sec:nofluxlif} that on $B_3 = \mathbb P^1 \times \mathbb P^2$ this mechanism does not lead to the generation of a superpotential by D3-brane instantons.

\subsection{General mechanism} \label{subsec_genflux}

The possibility that instanton flux can lift some of the deformation modes of a D3-brane instanton in Type IIB/F-theory  was pointed out  in \cite{Bianchi:2011qh}.
Consequently, the $\chi(\hD)=1$ condition is modified for instantons carrying instanton flux  \cite{Bianchi:2011qh}. 
However, instanton flux can only lift (some of the) zero-modes counted by $H^2(D, {\cal O}_D)$ or $H^2(D, K_{B_3}|_D)$ (see Table \ref{tab:instmap-main}), and not the other types of modes.

While in principle the flux lifting mechanism can be understood in full generality within F-theory  \cite{Bianchi:2011qh} (see \cite{Kerstan:2012cy} for the related M-theory formulation),
for practical purposes the simplest explanation is available in the weakly coupled Type IIB limit, provided it can be taken.  
Admissable instanton flux is then of the form \cite{Grimm:2011dj}
\be
{\cal F} \in H^{1,1}_-(\tilde D) \,.
\ee
In order for such flux to be able to lift some of the deformation modes, it must not lie in the space $\iota^*H^{1,1}_-(X_3)$ where $\iota: \tilde D \hookrightarrow  X_3$
is the inclusion of the instanton divisor $\tilde D$ into $X_3$.
Furthermore, the flux must be writable as \cite{Bianchi:2011qh}
\be \label{FasCi}
{\cal F} = {\cal F}_0 + \sum_a (C_a - C_a') \,.
\ee
Here ${\cal F}_0$ is chosen such as to satisfy the Freed-Witten anomaly along $\tilde D$ and $C_a$ is a rigid curve on $X_3$ with orientifold image $C_a'$.
As long as $C_a - C_a'$ is non-trivial as an element of $H^{1,1}_-(\tilde D)$ such flux forces the divisor $\tilde D$ onto the locus where it contains the rigid curves $C_a$ and $C_a'$, thereby lifting (some of) its deformation modes.

\subsection{No flux lifting for $B_3 = \mathbb P^1 \times \mathbb P^2$}
\label{sec:nofluxlif}

Consider now F-theory on an elliptic fibration $Y_4$ over the base $B_3 = \mathbb P^1 \times \mathbb P^2$.
This is a special case of the more general class of base spaces given by a $\mathbb P^1$-fibration $p: B_3 \to B_2$ (see the discussion around (\ref{pproj}) in Appendix \ref{sec:stringhet} for more details).
The trivial fibration $\mathbb P^1 \times \mathbb P^2$ has a section $S_-$ and we can furthermore consider the pullback $p^\ast(H)$ of the hyperplane class $H$ on $\mathbb P^2$ to $B_3$.
The most general divisor on $B_3$ is hence of the form $D = a S_- + p^*(bH)$.

From the general expression for the arithmetic genus (\ref{agen}) one finds that for the pullback $\hat D$ of this divisor to $Y_4$:\footnote{This uses that for a smooth Weierstrass model with section $\sigma$,
$c_2(Y_4)  = 12 \sigma \pi^*c_1(B_3) + 11 (\pi^*c_1(B_3))^2 + \pi^*c_2(B_3)$.
Together with the intersection numbers 
$S_- \cdot_{B_3} S_- = - S_- \cdot_{B_3} c_1({\cal L})$ (cf. (\ref{intnumbersS-S+})) this implies $\chi(\hD) = -\frac{1}{2} \int_{B_3} D^2 c_1(B_3)$.} 
\be
\chi(\hat D = \pi^*D) = - b^2    \qquad \quad {\rm for } \qquad D = a S_- + p^*(bH) \,.
\ee
According to the discussion in Section \ref{sec:instgency},
there are therefore no unfluxed instantons along divisors with $\chi(\hD) =1$ which would straightforwardly contribute to the superpotential \cite{Witten:1996bn}.

It is important to note, however, that there are divisors whose additional zero-modes are deformation zero-modes, rather than Goldstino zero-modes counted by $h^{(1,0)}(\hD)$.
In particular, for unfluxed instantons, one computes, for example,
\bea \label{pHinstanton}
h^{(i,0)}(\hat D = \pi^\ast D) &=& (1,0,0,2)    \qquad {\rm for} \, \, D = p^*(H) \, \\
h^{(i,0)}(\hat D= \pi^\ast D) &=&(1,0,0,1)     \qquad {\rm for} \, \, D = S_- \,.
\eea  
%In fact, as for $S_-$ this is the structure of zero-modes for every base $B_2$ as long as the twist bundle ${\cal L}$ is trivial.

This raises the question if the additional deformation zero-modes counted by $h^{(3,0)}(\hD)$ for both types of divisors are lifted if the instanton carries suitable instanton flux, as in the example where $B_3 = \mathbb P^3$  \cite{Bianchi:2011qh} reviewed in Section \ref{sec_P3}.
However, we now argue that this is not the case.

As for $S_-$, this follows already from the arguments given in \cite{Bianchi:2011qh}. In particular, deformation zero-modes cannot be lifted whenever the normal bundle exact sequence for the divisor $D$ on $B_3$ splits holomorphically in the sense that $T_{B_3}|_{D} = T_{D} \oplus N_{D/B_3}$. This is indeed the case here because $D=S_-$ is identified with the second factor in the direct product $B_3 = \mathbb P^1 \times \mathbb P^2$.
To show that also the zero-modes of $ D = p^*(H)$ cannot be lifted by instanton flux is more involved.
Following \cite{Bianchi:2011qh} we approach this problem in the orientifold limit, which can always be taken for a generic Weierstrass model over $B_3$.

According to the general procedure \cite{Sen:1997gv}, the Calabi-Yau 3-fold $X_3$ of the Type IIB orientifold is constructed as a double cover of $B_3$ given by a generic hypersurface in the toric ambient space $X_4$ shown in Table \ref{tab:X4},
\be \label{tX3hyper}
X_3: \qquad \xi^2 = P_{4,6}(z_i, u_j) \,.
\ee
The orientifold action is 
\be \label{invol}
s: \xi \to - \xi \,.
\ee
This space is a genus-one fibration over $\mathbb P^2$ with the genus-one fiber represented as a hypersurface in $\mathbb P_{1,1,2}$ with homogenous coordinates $[z_1 : z_2 : \xi]$.
In particular, $z_1=0$ is the divisor associated with a bi-section, as analysed in detail in \cite{Morrison:2014era}.

\begin{table}[t!]
  \centering
  \begin{tabular}{|c|c|c|c|c|c|}
   \hline $z_1$ & $z_2$ & $\xi$   & $u_1$ & $u_2$ & $u_3$  \\ \hline  
    1       &     1     &   2       &  0        &     0     &  0   \\  \hline 
    0       &     t=0     &   3       &  1        &     1     &  1  \\   \hline 
  \end{tabular}
  \caption{\small Scaling relations for toric ambient space $X_4$ of Calabi-Yau double cover $X_3$ associated with $B_3 = \mathbb P^1 \times \mathbb P^2$.}
  \label{tab:X4}
\end{table} 

Consider first the double cover $\tilde D = \tilde S_-$ of the divisor $D= S_-$ on $B_3$. A generic member in this family is identified with the vanishing locus of the polynomial $c_1 z_1 + c_2 z_2 = 0$.
Such a divisor is a K3-surface, as is easily checked with the help of the adjunction formula. In fact $X_3$ admits, in addition to the genus-one fibration, a K3-fibration over a rational curve, and $\tilde S_-$ is precisely the fiber of this second type of fibration.
This implies that $\tilde S_-$ cannot contain any rigid curves. According to the discussion around (\ref{FasCi}) there exists therefore no instanton flux that can lift the deformation mode on $\tilde S_-$
(and hence on $S_-$), in agreement with the general argument above. 

The double cover $\tilde D = \tilde p^*(H)$ corresponds to a divisor $a_1 u_1 + a_2 u_2 + a_3 u_3 = 0$ on $X_3$.
%One can check that $h^{1,1}(\tilde D) = 30$, leaving ample room for instanton flux to be summed over. 
Again we claim that there exists no instanton flux of the form (\ref{FasCi}) required to lift the two deformation modes. 
To see this, we need to understand the structure of rigid curves on the genus-one fibration $X_3$. Any holomorphic curve on $X_3$ is either a fibral curve, an curve on the base $B_2$ (i.e. lying in the section) or a linear combination. The base $B_2 = \mathbb P^2$ does not contain any rigid curve. The general fiber is not rigid either. This leaves as the only source of rigidity fibral curves into which the genus-one fiber degenerates.
Indeed, as analysed in \cite{Morrison:2014era}, over isolated points $p_a$ on $B_2$, the genus-one fiber $F$ splits into two rigid homologous rational curves $C_a$ and $\tilde C_a$, $F \rightarrow C_a + \tilde C_a$ with $[C_a] = [\tilde C_a] = \frac{1}{2} [F]$ in $H_2(X_3, \mathbb Z)$.
This happens over such points $p_a$ where the hypersurface equation (\ref{tX3hyper}) degenerates as
\be
(\xi + f_a)(\xi -f_a) = g_a^2
\ee
for $f_a(z_i,u_j)$ and $g_a(z_i,u_j)$ suitable polynomials. The curves $C_a: (\xi + f_a) = g_a = 0$ and $\tilde C_a (-\xi + f_a) = g_a = 0$ are then homologous rational curves.
We note that they 
map to one another under the orientifold involution (\ref{invol}), i.e. $C_a' := s(C_a)  = \tilde C_a$.

Furthermore, $C_a$ and $C_a'$ are homologous not only in $X_3$, but also within $\tilde D$. This is clear because the bi-section intersects $C_a$ and $C_a'$ each in one point and the two intersection points are exchanged by a monodromy along a ramification divisor   on $B_2$, corresponding to the existence of a chain connecting $C_a$  and $C_a'$.
In the present case, the ramification divisor is in the class $2 \bar K_{B_2} = 6 H$,\footnote{In the notation of \cite{Mayrhofer:2014haa}, see Table 2.1 therein, the specific fibration corresponds to the choice $\beta = \bar K_{B_2}$, while from equ. (2.13) and Table 2.2 one infers that the ramification divisor is in class $[c_4] = 4\bar K_{B_2} - 2 \beta$.}
 and since 
 since this ramification locus of the bi-section always intersects the class $H$ on $\mathbb P^2$, it cannot be arranged for $C_a$ and $C_a'$ to be not homologous on $\tilde D$.
As a result, the zero-modes cannot be lifted by instanton flux of the form (\ref{FasCi}).

One can convince oneself that this problem persists for all other divisors on $B_3$. As a result, we have identified an example of an F-theory compactification with no divisors supporting fluxed or influxed instanton that can contribute to the superpotential without invoking other mechanisms of zero-mode lifting.

%%%%%%%%%%%%%%%%%%%%%%%%%%%%%%%%%%%%%%%%%%%%%%%%%%%%%%%%%%%%%%%%%%%%%%%%%%%%%
\section{Stringy heterotic instantons}
\label{sec:stringhet}

In this Appendix we study stringy NS5-brane instantons in the heterotic string. 
Their importance derives from the fact that they can in general induce a superpotential in heterotic string vacua even in situations where worldsheet instanton effects are known
to lead to no such correction. 
In the sequel we provide what we believe is the first example where a stringy NS5-brane instanton is guaranteed to contribute to the superpotential.

While the zero-modes of NS5-brane instantons are difficult to study directly in the heterotic frame, they can in suitable settings be understood by duality. As shown in Figure \ref{fig:instmap}, heterotic-Type I duality maps NS5 instantons to D5 instantons in Type I,
%The case where the D5 instantons carry the same bundle as the D9 branes, so $H=V$, leads to their interpretation as gauge instantons. While the case $H \neq V$ is termed stringy instantons. 
while under heterotic-Type IIB duality, NS5 instantons map to certain D3 instantons.
The latter can sometimes be interpreted as gauge instantons, but in general can only be understood as stringy instantons. 

In the sequel we will recall the dual description of heterotic NS5-brane instantons as D3-brane instantons in F-theory, first described in \cite{Witten:1996bn}. 
We will then provide an example where the latter instanton is guaranteed to contribute to the superpotential; this exemplifies that stringy NS5-brane instantons can indeed correct the superpotential even in absence of confining gauge symmetry or other types of gauge instantons in heterotic string theory.

The heterotic string compactified on an elliptically fibered Calabi-Yau 3-fold $Z_3$ with base $B_2$ is dual to F-theory on an elliptic 4-fold $Y_4$ whose base 
$B_3$  admits a fibration \cite{Friedman:1997yq}
\bea \label{pproj}
p :\quad \mathbb{P}^1 \ \rightarrow & \  \ B_{3} \cr 
& \ \ \downarrow \cr 
& \ \  B_2 
\eea
Such $B_3$ can be expressed as the projectivised bundle
\be
B_3 = \mathbb P({\cal O}_{B_2}  \oplus {\cal L}) \,,
\ee
where the line bundle ${\cal L}$ on $B_2$ describes the twist of the $\mathbb P^1$-fibration. 
There are two distinguished sections, whose associated divisor classes $S_-$ and $S_+$ are given by
\be
S_-  = c_1({\cal O}(1)) \,, \qquad \quad S_+  =S_- + c_1({\cal L}) \,,
\ee
and which satisfy
\be \label{intnumbersS-S+}
S_- \cdot_{B_3} S_+ = 0 \,.
\ee
Here ${\cal O}(1)$ denotes the line bundle on $B_3$ whose restriction to the $\mathbb P^1$-fiber is the line bundle ${\cal O}(1)$.
In terms of these quantities,
\be \label{c1B3fibation}
c_1(B_3) = 2 S_- + p^*(c_1({\cal L}) +c_1(B_2)) \,.
\ee
The heterotic gauge group in the two $E_8$ factors maps to the gauge groups of the 7-brane stacks localised along $S_-$ and $S_+$, respectively. For more details we refer e.g. to \cite{Anderson:2014gla} and references therein.

Consider an M5-brane instanton on $Y_4$ wrapping a divisor $\hat D = \pi^*(D)$ for $D$ a divisor on $B_3$.
Such instantons dualize to the following objects in the heterotic frame \cite{Witten:1996bn}: \vspace{1mm}
\begin{center}
\begin{minipage}{3cm}
$D = S_\pm$      \\
$D = p^*(C)$   
\end{minipage}
\begin{minipage}{1cm}
 $\Longleftrightarrow$
\end{minipage}
\begin{minipage}{6cm}
NS5-brane instanton on $Z_3$ \\
worldsheet-instanton on $C \subset B_2$
\end{minipage}
\end{center}
Here $C$ is a curve on  $B_2$. %, which is the common base to the $\mathbb P^1$-fibration $B_3$ and the heterotic elliptic fibration $Z_3$.
An instanton along a general divisor in class $D = a S_- + p^*(C)$ then maps to a bound state of an NS5-worldsheet instanton. Note that for all types of instantons one must again sum over all possible instanton flux configurations.
For a recent systematic study of D3-brane instantons in the context of F-theory/heterotic duality, see also \cite{Anderson:2015yzz}.

After this review we now exemplify that stringy NS5-brane instantons can indeed contribute to the superpotential.
For the purpose of providing such an example it suffices to specialise to $B_2 = \mathbb P^2$. The possible choices of twist bundle are simply ${\cal L}= {\cal O}_{\mathbb P^2}(t H)$ with $H$ the hyperplane class of $\mathbb P^2$. 
As analysed in \cite{Anderson:2014gla},
for $t \geq 4$, the theory contains a non-Higgsable gauge group along the section $S_-$. In particular, this invalides the expression in the fourth column of Table \ref{tab:instmap-main}, which assumed that $\hD$ is a smooth divisor.
For $t=0,1,2,3$, on the other hand, we can take the M-theory Calabi-Yau 4-fold $Y_4$ to be a smooth Weierstrass model over $B_3$, leading to a  4d $N=1$ theory with trivial gauge group. 
For simplicity we focus on such smooth situations. For unfluxed instantons along $S_\mp$, evaluating the multiplicities in Table \ref{tab:instmap-main} gives\footnote{Eq. (\ref{c1B3fibation}) and (\ref{intnumbersS-S+}) imply that $K_{B_3}|_{S_-} \simeq {\cal O}_{\mathbb P^2}(t-3)$ and $K_{B_3}|_{S_+} \simeq {\cal O}_{\mathbb P^2}(-t-3)$. The result then follows from Bott's theorem for line bundle cohomologies on $\mathbb P^2$.}
\bea
&t=0:       &h^{(i,0)}(\pi^*S_-) =(1,0,0,1) \,, \qquad h^{(i,0)}(\pi^* S_+) =(1,0,0,1)   \label{zeromodesS-t0} \\
&t=1,2:       &h^{(i,0)}(\pi^* S_-) =(1,0,0,0) \,, \qquad h^{(i,0)}(\pi^* S_+) =(1,0,0,a) \\
&t= 3:       &h^{(i,0)}(\pi^* S_-) =(1,b,0,0) \,, \qquad h^{(i,0)}(\pi^* S_+) =(1,0,0,a) 
%&t=0:       &h^{(i,0)}(S_-) =(1,0,0,1) \,, \qquad h^{(i,0)}(S_+) =(1,0,0,1) \\
\eea
for
\be
a = \frac{1}{2}(t+2)(t+1) \,, \qquad \quad b= \frac12  (t-1)(t-2) \,.
\ee
For $t=1,2$, we conclude that an unfluxed NS5-brane instanton in the heterotic theory dual to the instanton along $S_-$ straightforwardly contributes to the superpotential, even without invoking other effects.
%This already constitutes the first main result of this analysis. 
For $t=0$ and $t=3$, at least one of the unfluxed instantons along $S_-$ or $S_+$ has only extra deformation modes in $H^{3,0}(\hD)$.\footnote{The statement for $t=0$ holds for any base $B_2$ and trivial twisting $c_1({\cal L}) =0$.} 
Such zero-modes are {\it in principle} amenable to lifting by instanton flux \cite{Bianchi:2011qh}, though whether or not this is possible depends on the intricate details of the geometry, and to the lifting mechanism involving spacetime filling D3-branes studied in section \ref{sec_P1P2}.

%%%%%%%%%%%%%%%%%%%%%%%%%%%%%%%%%%%%%%%%%%%%%%%%%%%%%%%%%%%%%%%%%%%%%%%%%%%%%
\section{Superpotentials in heterotic compactifications}
\label{sec:hetvac}

In this section we study superpotentials in the heterotic string. The $E_8 \times E_8$ or $SO(32)$ heterotic string on a Calabi-Yau threefold gives an $\cN=1$ supergravity in four dimensions. Since the heterotic string in ten dimensions includes a non-Abelian gauge sector, before worrying about instanton contributions to the superpotential we should consider gauge theoretic contributions such as gaugino condensation.\footnote{In some sense, this applies also to Type II string theories, since introducing orientifold planes to break the supersymmetry to $\cN=1$ requires also introducing branes to cancel the tadpoles.} With regards to instantons, there are three types of instantons which can potentially contribute to the superpotential: worldsheet instantons, NS5 gauge instantons, and NS5 stringy instantons.\footnote{In the presence of NS5 spacetime-filling branes, there are also E-string instantons.} 

NS5 stringy instantons are difficult to study explicitly in the heterotic context.
As exemplified in Appendix \ref{sec:stringhet} they can in general contribute to the superpotential as stringy instantons. Unless this can be ruled out in specific settings, this possibility
 therefore generally invalidates any candidate for a 
theory with $W=0$.

The aim of this Appendix is to determine whether it is possible to show that there are no heterotic compactifications with vanishing superpotential even if we leave aside such stringy NS5 brane instantons. 
We will not arrive at a definitive conclusion, due to recent developments in \cite{Buchbinder:2019hyb}, but will present a serious contender for a setup with no gauge or worldsheet instanton contributions.
We therefore believe that a claim for the universal  generation of a superpotential in generic heterotic compactifications must involve stringy NS5 instantons. 

Before proceeding let us note that Type I vacua are S-dual to heterotic vacua and so are in this sense also covered by the analysis of this section. See, however, e.g. \cite{Camara:2007dy,Camara:2008zk,Camara:2010zm} for studies of instantons directly in a Type I setting.

Heterotic worldsheet instantons  are dual to certain D3 instantons as reviewed in Appendix \ref{sec:stringhet}. While they generically generate a superpotential, there are some special circumstances where this can be shown not to occur  \cite{Distler:1986wm,Silverstein:1995re,Beasley:2003fx} (see \cite{Buchbinder:2016rmw,Buchbinder:2017azb,Buchbinder:2019hyb,Buchbinder:2019eal} for recent work). 
%One setup is where appropriate line-bundles are turned on such that they have a non-trivial restriction to every holomorphic curve \cite{Distler:1986wm,Buchbinder:2019eal}. More generally, 
It was proposed that worldsheet instantons are absent on favourable complete-intersection Calabi-Yau manifolds embedded in weighted projective spaces or toric spaces, and for which the bundle is inherited from the ambient space \cite{Beasley:2003fx,Buchbinder:2016rmw}. This Beasley-Witten (BW) theorem is based on earlier work in \cite{Silverstein:1995re} which showed that worldsheet instantons do not destabilise certain vacua which have a $(0,2)$ sigma model description. The argument relies on compactifying the moduli space as in the Riemann sphere. The superpotential is then a holomorphic function on the compact moduli space, and it has a universal zero at the decompactification limit. It must therefore either be exactly vanishing or have somewhere a pole. To determine this one needs to know the full moduli space, which includes small volumes away from the supergravity regime. A sigma-model description allows for this, and can be used to argue for the absence of poles. Recently, in \cite{Buchbinder:2019hyb,Buchbinder:2019eal} it was shown that many compactifications which apparently satisfy the requirements for the BW theorem in fact do have non-vanishing worldsheet instantons. This was attributed to the non-compactness of the bundle moduli space, and therefore suggests that actually this must be imposed as a further independent criterion which is difficult to check. If there is a sigma model description of the setting, then in \cite{Bertolini:2014dna} a prescription for checking this compactness was given.

Let us now discuss the consequences of the Beasley-Witten theorem for standard embeddings and for non-standard embeddings in turn.

\subsection{Standard embeddings}

In the standard embedding the gauge bundle is identified with the tangent bundle of the Calabi-Yau.
%In this case the resulting $\cN=1$ vacuum has no classical potential.
For favorable Calabi-Yau spaces, the gauge bundle is therefore inherited from the ambient space, and the standard embedding  automatically satisfies the conditions for the Beasley-Witten theorem to hold. 
Furthermore, unlike for more general gauge bundles, the standard embedding does not require the introduction of spacetime-filling NS5-branes. These objects are dual to D3-branes in F-theory which, as argued in Section \ref{sec_P1P2}, affect the dynamics of instantons in a non-trivial way.
Both these points make this class of constructions an interesting testing ground for the generation of a superpotential.
Prior to discussing the worldsheet instantons, however, we must analyze the potential generation of a superpotential by strong gauge dynamics.

The results of the following discussion can be summarized as follows:
\begin{enumerate}
\item For the standard embedding, there is always a gaugino condensate generating a superpotential $W_{\rm gauge}$. This superpotential alone may or may not allow for a solution to $W_{\rm gauge} = dW_{\rm gauge} =0$, even though genericity arguments would suggest this not to occur except on special backgrounds.
\item
If $\chi(Z_3) \neq 0$ with $Z_3$ the heterotic Calabi-Yau, the Beasley-Witten vanishing theorem for the superpotential due to worldsheet instantons is a priori not valid even on favorable manifolds, while we are not aware of favorable examples for $\chi(Z_3) = 0$ which do not lead to $\cN \geq 2$ supersymmetry.
\end{enumerate}

To see the first point, recall that in the $E_8 \times E_8$ heterotic string the gauge bundle is completely embedded into only one of the $E_8$ factors. This results in a non-perturbative potential induced through gaugino condensation in the other $E_8$ factor, which has no massless charged matter.
The standard embedding for the $SO(32)$ (or more precisely $Spin(32)/\mathbb Z_2$) heterotic string breaks the gauge group as
\bea
SO(32) &\rightarrow& SO(26) \times U(1)  \;.
\eea
This leads to charged chiral matter under the $SO(26)$ non-Abelian factor which can prevent gaugino condensation. The chirality of the charged matter in representation ${\bf R}$, denoted $\chi\left(\bf R \right)$, is set by the Euler number of the Calabi-Yau $\chi\left(Z_3\right)$,
\bea
\chi(({\bf 26})_1) &\equiv& \#{\bf 26}_1 - \#{\bf \overline{26}}_{1} =  \chi(Z_3) \;, \nn \\
\chi(({\bf 1})_2) &\equiv& \#{\bf 1}_2 - \#{\bf 1}_{-2} =  - \chi(Z_3) \;,
\eea
where the subscript denotes the $U(1)$ charge. The chiral matter is protected and hence massless, and therefore for sufficiently large $\chi\left(Z_3\right)$ will prevent gaugino condensation in the infrared. However, there is also a 1-loop D-term which takes the form (see, for example \cite{Witten:2013cia,})
\be
D_{\chi} \sim \sum_{i}( |\phi_i|^2 - |\tilde\phi_i|^2 )     + 2 \,  (\sum_{j} |\psi_j|^2 - |\tilde\psi_j|^2) + \frac{c}{2} g_s^2  \,  \chi(Z_3) \,,
\label{hetdter}
\ee
where $\phi_i$ and $\psi_j$ denote the complex scalars in the $\cN=1$ chiral multiplets in representation $({\bf 26})_1$ and $({\bf 1})_2$ and $\tilde\phi_i$ and $\tilde\psi_j$ the complex scalars in representation $({\bf 26})_{-1}$ and $({\bf 1})_{-2}$. The constant $c$ is of order one, and $g_s$ is the string coupling. This D-term implies that the $U(1)$ must be broken in the vacuum precisely when $\chi(Z_3)\neq 0$, i.e. when there is potentially no gaugino condensation. The breaking of the $U(1)$ then removes the net chirality. From the field theory perspective we can look at the superpotential operators
\be
W \supset {\bf 1}_{-2} {\bf 26}_1 {\bf 26}_1 + \mathrm{c.c.} \;.
\label{hetsupyuk}
\ee
From (\ref{hetdter}) we see that if, say, $ \chi(Z_3) > 0$, then we need to give an expectation value to the $({\bf 1})_{-2}$ fields which gives a mass to any chiral net number of ${\bf 26}_1$ fields. That there always exists an appropriate $({\bf 1})_{-2}$ field to give an expectation value to was shown in \cite{AtickDixonSen} for a wide class of models, and conjectured to hold generally in \cite{Aldazabal:2018nsj}. The resulting vacuum then restores supersymmetry at the perturbation theory level (see, for example \cite{Pius:2014gza,Sen:2015uoa} for recent studies). Note that the superpotential (\ref{hetsupyuk}) also shows that there is an obstruction to solving the D-term condition by giving an expectation value to the ${\bf 26}_1$ fields and thereby also breaking the non-Abelian gauge group.

We conclude that no matter the value of $\chi(Z_3)$, the spectrum of the standard embedding for the $SO(32)$ heterotic string is non-chiral with a non-Abelian gauge group. This means that at a generic point in moduli space all the matter will be massive, leading to a non-perturbative superpotential in the infrared. More precisely, there exists a non-perturbative superpotential with a pre-factor which is moduli dependent and vanishes on certain special sub-loci of the moduli space where a sufficient number of vector-like charged states are massless. If we denote the moduli fields (bundle and complex-structure) collectively as $u^i$, then the superpotential takes the schematic form
\be
W_{\rm gauge} \sim f\left( u^i \right) e^{-\frac{a}{g_s} } \;.
\ee
Here $a$ is some constant and $f$ is a function of the moduli which vanishes on certain special loci denoted $u^i = u^i_0$, i.e.  $f\left( u^i_0 \right) =0$. 
We cannot exclude in full generality that on the locus $u^i_0$ also $dW_{\rm gauge}  = 0$, though based on genericity arguments alone this is not expected; in any event, $W_{\rm gauge}$  does not include the remaining contributions to the superpotential, especially from stringy NS5-brane instantons.\footnote{This is in notable difference to the non-generic setting in Section \ref{sec:N=1orbvacu}, where we do argue for the existence of a sub-locus in moduli space with $W = dW=0$.}

Having studied the effect of gaugino condensation we can now turn to the generation of a superpotential via 
 worldsheet instantons. In the case $\chi(Z_3) \neq 0$ we have just seen that the D-term requires an expectation value for a charged field. This corresponds to a deformation of the gauge bundle, and the deformed  bundle is no longer expected to be inherited from the ambient space. Therefore, the BW vanishing theorem will not hold in general. In the case $\chi(Z_3)=0$, the BW theorem should hold at least as long as the manifold is a favourable embedding in a toric space.
  
Let us consider the Complete Intersection Calabi-Yau manifolds (CICYs) as a sample set. There are 52 cases with $\chi(Z_3) = 0$. They are composed of 22 cases of direct product manifolds with a torus (which have enhanced supersymmetry), 15 cases with $h^{1,1}=19$ which are representations of the Schoen manifold, and 15 cases with $h^{1,1}=15$, which are also the same manifold. The two non-product manifolds are not favourable when embedded in the projective spaces.\footnote{However, in \cite{Anderson:2017aux} it was argued that one can utilise an ineffective splitting to rewrite the Schoen manifold as a favourable embedding in a certain del Pezzo space, and the second manifold as a favourable embedding in projective spaces. Then an interesting question is whether the BW theorem can hold when the manifolds are rewritten this way. One piece of evidence against it could be that the Schoen manifold does support bundles which admit worldsheet instanton corrections \cite{Braun:2018fdp}.} Interestingly, this leaves as the only class of CICYs
with $\chi(Z_3) =0$ those which lead to enhanced supersymmetry, which are therefore not of relevance when it comes to studying the generation of a superpotential.

\subsection{Non-standard embeddings}

We have accumulated evidence (though not provided a definitive proof) that standard embeddings for the heterotic string lead to the generation of a superpotential. It is also possible to consider bundles which are not the standard embedding. Indeed, the largest set of realistic heterotic models in the geometric regime are constructed this way \cite{Anderson:2011ns,Anderson:2012yf,Anderson:2013xka}. When studying non-standard embeddings it is crucial for our purposes to consider the Bianchi identity, or tadpole condition, which takes the schematic form
\be
\mathrm{Tr}\left( F \wedge F \right) - \mathrm{Tr}\left( R \wedge R \right) = \delta_{\mathrm{NS5}} \;.
\label{BIhet}
\ee
The first term is the contribution from the second Chern class of the bundle, the second term from the curvature, and if these are not equal then they must be balanced by the inclusion of spacetime filling NS5-branes.  
In this case it is no longer clear that the BW vanishing theorem holds. Indeed, dualising  the results of Section \ref{sec_P1P2} to the heterotic string suggests that there are new potential instanton contributions in the presence of such spacetime-filling NS5 branes. %$Furthermore, if $c_1(V) \neq 0$, then a non-zero restriction of this class of the worldsheet leads to additional charged zero-modes which again modify the computation of the instanton effect.
The best controlled candidates for theories with $W=0$ are therefore embeddings which saturate the Bianchi identity and so do not require any NS5 branes.

The most natural way to saturate the Bianchi identity (\ref{BIhet}) in the $E_8 \times E_8$ heterotic string is through the introduction of a further bundle  on the hidden $E_8$. We will consider two types of models which will illustrate two aspects of the vacuum stability. 

The first model is based on the case studied in \cite{Nibbelink:2015ixa}.\footnote{We restrict to the case $k=0$ in the notation of \cite{Nibbelink:2015ixa}.} This is a compactification utilising line bundles, based on the earlier work \cite{Blumenhagen:2005ga,Anderson:2011ns,Anderson:2012yf,Anderson:2013xka}. There are line bundles on both of the $E_8$ factors, such that the Bianchi identity is saturated and there is no need to introduce NS5 branes. The resulting gauge group is $SU(5)\times SU(4) \times U(1)^9$, and the matter spectrum is such that there is no gaugino condensation or gauge instantons in the non-Abelian factors. This is a good testing ground to see if the D-terms require breaking the $U(1)$ symmetries in such vacuum. While in \cite{Nibbelink:2015ixa} the D-terms were solved with singlet expectation values which break the $U(1)$s, this need not be the case in general. Indeed, we find the following solution for the divisor volumes ${\cal V}_i$, with $i=1,2,3,4$,
\be
{\cal V}_2 = {\cal V}_1 + 6 X \;,\;\; {\cal V}_3 = {\cal V}_1 + 2 X \;,\;\; {\cal V}_4 = {\cal V}_1 + 8 X \;. 
\ee
Here $X$ denotes the 1-loop contribution to the D-term, as in the last term of (\ref{hetdter}). The important point is that the 1-loop contribution is cancelled by the relative differences in the volumes of the divisors, so the latter can remain large and in the perturbative regime. This proves that there are examples where there are no worldsheet instantons, no gaugino condensation in the infrared, and the $U(1)$s need not be broken by the D-terms. There is still a subtlety in this model however, because if appropriate F-flat directions exist in the charged singlet moduli space, then it is possible (even if not enforced by the D-terms) to break all the $U(1)$s and make the spectrum non-chiral with respect to the $SU(4)$. This would then imply a gaugino condensation contribution to the superpotential which vanishes on the split locus where the charged singlets have vanishing expectation values. However, this does not guarantee that this locus is the minimum of the potential, and so the vacuum may still be destabilised due to gaugino condensation. 

To make things even sharper we can consider a second model which is purely non-Abelian. We consider compactifications of the $E_8 \times E_8$ heterotic string with monad bundles studied in \cite{Anderson:2007nc}. First note that monad bundles are expected to satisfy the BW vanishing theorem, at least on favourable Calabi-Yau manifolds \cite{Beasley:2003fx}. We will consider the quintic, which is favourable, and so there should not be any worldsheet instanton corrections. From Table 4 in \cite{Anderson:2007nc} we see that there exist two $SU(3)$ monad bundles with second Chern classes of 3 and 7 (relative to the square of the overall K\"ahler form). Together with $c_2(Z_3) = 10$ for the quintic,  herefore taking one such bundle on each $E_8$ factor saturates the Bianchi identity and there are no NS5 branes. The resulting theory has gauge group $E_6 \times E_6$, with 60 chiral ${\bf 27}$s under the first $E_6$ and 15 under the second. On general grounds it is expected that there is no gaugino condensation for number of ${\bf 27}$s larger than 2, which is easily satisfied for both the $E_6$ factors. There are no $U(1)$s, and so no D-terms, and it is not possible to lift the chiral matter without also breaking the non-Abelian gauge group. 

This model suggests, as claimed at the start of this Appendix, that the only potential universal obstruction to compactifications with $W=0$ in the heterotic string must come from stringy NS5 instantons. The only subtlety is the one discussed in the recent work \cite{Buchbinder:2019hyb,Buchbinder:2019eal}, which showed that sometimes the BW theorem does not hold due to non-compactness of the moduli space. It would be interesting to apply the more refined criteria of \cite{Buchbinder:2019hyb,Buchbinder:2019eal} to this, and other similar, examples.

\section{Type IIA, Type I and M-theory vacua} \label{App_IIA}

Type IIA compactifications on Calabi-Yau orientifolds, and M-theory compactifications on $G_2$ manifolds, lead to $\cN=1$ four-dimensional  theories. In these cases the geometry associated to the relevant instantons is real, rather than holomorphic, specifically special Lagrangian 3-cycles for IIA D2-brane instantons and associative manifolds for membrane instantons on $G_2$ manifolds. It is difficult to study such real submanifolds explicitly, and in particular to determine the associated zero-modes. Things are complicated further by an incomplete understanding of the effects of worldvolume fluxes in Type IIA, and the dual higher derivative effects in M-theory.

Type IIA vacua can be grouped into the F-theory vacua of section \ref{sec:ftvac} since they are related through mirror symmetry. This is only the case for one of two types of orientifold projections in IIA, the other case being mapped to a Type I mirror. The Type I vacua are dual to heterotic $SO(32)$ vacua. For the standard embedding of the gauge bundle in such vacua, we argue in appendix \ref{sec:hetvac} that there is a non-perturbative superpotential induced due to gaugino condensation. The superpotential is therefore not vanishing everywhere in field space, and it is difficult to determine if there is a minimum where the superpotential vanishes. More general gauge bundle choices, different from the standard embedding, are difficult to study at the non-perturbative level, especially in the presence of spacetime-filling NS5-branes which can affect the worldsheet instanton corrections on the heterotic side (and which map to D1 instantons in Type I). Overall, we do not find the Type IIA and Type I settings offer sufficiently good control to search for examples with vanishing corrections. 

The case of M-theory on $G_2$ manifolds, which yields an $\cN=1$ supergravity \cite{Papadopoulos:1995da}, receives potential instanton contributions to the superpotential from M2 branes on associative 3-cycles. Since the microscopic theory for M2 branes is not known, it is not possible to calculate the condition for contributing to the superpotential explicitly as in Type II. But it is natural to perform an analysis similar to \cite{Witten:1996bn} in terms of deformation modes of the cycle. This was performed in \cite{Harvey:1999as} which led to the proposal that M2 instantons contribute to the superpotential if they wrap rigid associative 3-cycles. See also \cite{Acharya:1998pm} for similar early work. The topic of instantons on $G_2$ manifolds is actively studied, boosted by the new constructions of $G_2$  manifolds using Twisted Connected Sums (TCS) \cite{Kovalev_2003,KOVALEV_2011,Corti_2013} (see \cite{joyce1996} for the earlier resolved orbifold constructions). We refer, in particular, to \cite{Corti_2015} for studies of associative cycles on such manifolds, as well as \cite{menet2015construction,S_Earp_20152,S_Earp_2015,Walpuski_2013,Walpuski_2016,Walpuski_2017,Joyce_2018} for more general studies of associative cycles and instantons. The construction of TCS $G_2$  manifolds lends itself nicely to duality with heterotic and Type II string theories, which has been studied in \cite{Halverson:2015vta,Braun:2017uku,Braun:2018fdp,Acharya:2018nbo,Braun:2019wnj}.

Studying instantons in M-theory has some unique elements. For example, because all the massless fields have axionic components, the Pfaffian prefactor of the instanton must be either a constant or a modular function of the fields. There are few known cases of the construction of such modular functions for $G_2$ manifolds, one example being in \cite{Braun:2018fdp}. This example, the so-called $E_8(\times E_8)$ superpotential, also has duals in the heterotic string and in F-theory, and so is discussed in section \ref{sec:ftvac}. Another unique aspect is that all the instantons come from a single type of brane, the M2, which also does not support worldvolume fluxes. This would suggest that $G_2$ manifolds can be good examples of vacua which receive no corrections if the $G_2$ manifold supports no rigid associative cycles. However, there are no such known examples (to our knowledge).\footnote{The closest constructions we found are the TCS geometries in \cite{Corti_2015} (Table 5) which have no rigid associative cycles that are constructed through a particular methodology, leaving open the possibility that other constructions of associative cycles on these manifolds might be possible.} More generally, the analogues of effects that can lift zero-modes in Type II constructions, such as fluxes, are poorly understood in the $G_2$ setting. Possibly this can be attributed to higher fermions operators or higher derivative terms, such as the quartic fermionic terms analyzed in \cite{Beasley:2003fx}.

It is worth noting that mathematically, $G_2$ vacua which have no instanton corrections are denoted as unobstructed $G_2$ manifolds in \cite{Joyce_2018}. There are simple examples of unobstructed $G_2$ manifolds given in \cite{Joyce_2018}, but these actually preserve $\cN=2$ supersymmetry. 

%%%%%%%%%%%%%%%%%%%%%%%%%%%%%%%%%%%%%%%%%%%%%%%%%%%%%%%%%%%%%%%%%
\subsubsection*{Flux vacua}
\label{sec:flvac}

Turning on background closed-string fluxes will in general induce a classical superpotential \cite{Gukov:1999ya}. However, there may still exist vacua where this superpotential vanishes. For example, in Type IIB string theory there are points in complex-structure moduli space where the flux superpotential vanishes \cite{DeWolfe:2004ns,Dine:2005gz,Palti:2007pm}. However, this does not directly affect the presence or absence of instantons, which in this case would be in the K\"ahler moduli. If anything, Type IIB background fluxes might lift D3-brane instanton zero-modes (see \cite{Bianchi:2012kt} for the state of the art and further references), allowing them to contribute to the superpotential. 

One aspect of fluxes which does forbid certain instantons are anomalies induced on the D-brane worldvolume. For example, in Type IIA string theory turning on NS H-flux through a 3-cycle forbids any D2 instantons to occur on that cycle through a Freed-Witten anomaly. This can also be understood from a supergravity perspective in terms of the flux gauging an isometry in the field space which must be an exact flat direction \cite{KashaniPoor:2005si}. However, there is no known example where there is a vacuum of vanishing superpotential where all the instantons are projected out this way.\footnote{More generally, this is part of an interaction between instantons and twisted K-theory, see for example \cite{Palti:2019pca} for a discussion of this.}

\bibliographystyle{jhep}
\bibliography{susyswamp.bib}  
%$\bibliographystyle{custom1}
\end{document}